# Conceptual Modeling: Topics, Themes, and Technology Trends

Veda C. Storey

Computer Information Systems, J. Mack Robinson College of Business, Georgia State University, USA 30302, vstorey@gsu.edu

Roman Lukyanenko

McIntire School of Commerce, University of Virginia, VA 22903, romanl@virginia.edu

Arturo Castellanos

Raymond A. Mason School of Business, William & Mary, Williamsburg, VA 23185, aacastellanosb@wm.edu

Conceptual modeling is an important part of information systems development and use that involves identifying and representing relevant aspects of reality. Although the past decades have experienced continuous digitalization of services and products that impact business and society, conceptual modeling efforts are still required to support new technologies as they emerge. This paper surveys research on conceptual modeling over the past five decades and shows how its topics and trends continue to evolve to accommodate emerging technologies, while remaining grounded in basic constructs. We survey over 5,300 papers that address conceptual modeling topics from the 1970s to the present, which are collected from 35 multidisciplinary journals and conferences, and use them as the basis from which to analyze the progression of conceptual modeling. The important role that conceptual modeling should play in our evolving digital world is discussed, and future research directions proposed.

## 1 INTRODUCTION

As human society deepens its reliance on information systems and information technology, the need to develop information systems in an efficient, rigorous, and dependable way, is vital. Conceptual modeling has long been recognized as a valuable foundation from which to develop information systems because it involves extracting concepts from the real-world, or application domain, and representing them in a way that supports interaction between designers and users of the systems. We define conceptual modeling as: *an activity that occurs during information systems development and use that involves capturing, abstracting, and representing relevant aspects of reality, to support understanding, communication, design, and decision making*. Conceptual models are comprised of constructs, such as entities, events, goals, attributes, relationships, roles, and processes, connected by well-defined rules.

The field of conceptual modeling emerged in the 1970s. It was initially understood as a phase of information systems development that systematically captures user requirements for database design. For example, relational database design, especially in large organizations, was preceded by conceptual modeling using entity-relationship diagrams (ERD), extended entity-relationship diagrams (EERD) [4, 177, 212, 222] or class diagrams in the Unified Modeling Language (UML) [51]. Over time, the field has expanded and matured, making contributions to requirements engineering, knowledge representation, process modeling, goal and value representation, ontology development, philosophy, and more recently, data analytics [54, 55, 87, 146, 148]. In the field of process modeling and process engineering, for example, popular conceptual modeling languages include Business Process Modeling and Notation (BPMN), Data Flow Diagrams (DFD), activity diagrams in UML, Event-driven process chains (EPC), and others. Some languages are designed for specific applications, ranging from wide-domain applicability, such as enterprise modeling (e.g., ArchiMate) to more niche languages, such as Formalized Administrative Notation (FAN) originally designed to support administrative workers in Argentina [12, 103].

Because conceptual modeling activities include abstraction and representation, they provide multiple benefits for domain understanding and comprehension [30, 226]. These models help structure reality by omitting aspects of the domain deemed irrelevant for some purpose [167]. Conceptual models reduce the complexity of the domains to be represented and help cope with the complexity of the information systems development process [51, 76, 202, 214]. As such, conceptual models may aid in decision making and problem solving [242].

Conceptual models are also commonly viewed as boundary objects [20]. A boundary object is "an artifact or a concept with enough structure to support activities within separate social worlds, and enough elasticity to cut across multiple social worlds" [211]. Indeed, conceptual models are frequently used by both technical and non-technical business users to gain a common basis for understanding the goals and requirements of systems to be built or the data to be used in decision making.

Conceptual modeling as a recognized field is now entering its 6th decade of practice. The proven record of conceptual modeling in the context of information technology development and use suggests it is posed to remain an important development and analysis tool for the foreseeable future. At the same time, to remain relevant, conceptual modeling needs to continue evolving and adapting to new technologies and application domains, as well as continued digitalization [127]. To ensure the future impact and effective use of conceptual modeling, it is valuable to analyze what has been accomplished to date, and to identify fruitful areas for future development as is the goal of this survey.

To guide the evolution of conceptual modeling research, various frameworks have been developed [131, 146, 158, 184, 196, 236] where the authors consider the state of the art in theory and practice. Most of these efforts do not engage in a comprehensive survey of conceptual modeling publications. Instead, they are typically based on new conceptual modeling assumptions (e.g., mediation versus representation) and provide guidance on their applications to the conceptual modeling community. These frameworks lack the nuances of a comprehensive, structured literature review, aimed at inclusive coverage of the topics at granular and high-abstraction levels [171]. A comprehensive and structured literature review promises fruitful ideas for specific research projects in conceptual modeling and the identification of broader trends and long-term research opportunities.

Coinciding with the 50-year anniversary of conceptual modeling, the objectives of this paper are to: review concepts, topics, and themes, as research in conceptual modeling has progressed over time; and identify needed continued research due to on-going technology advances and increased digital ubiquity. The contributions are to: provide a review of conceptual modeling; show how research topics related to conceptual modeling have evolved; and propose how research on conceptual modeling can continue to contribute to our digital society.

We reviewed over 5,300 papers from 35 related journals and conferences which, to the best of our knowledge, is the largest corpus of conceptual modeling research ever analyzed. These publications were retrieved from different, relevant disciplines including computer science, information systems, software engineering, human-computer interaction, and database design. We cataloged the papers into a text corpus and analyzed them with a mixed method approach. Our results are based on natural language processing (using LDA analysis of full-text papers). We also developed a specialized *conceptual modeling language model* to identify semantically similar terms and a topic model to identify meaningful groups of papers whose topics are closely related. Finally, we augmented the quantitative evidence with qualitative analysis and insights. This is an effort to review broadly the conceptual modeling literature belonging to different disciplines, genres, and levels of maturity (e.g., journals with multi-year review cycles, short conference papers, and long conference papers). Following this broad survey, future research directions are identified for how conceptual modeling can, and should, evolve.

This paper proceeds as follows. Section 2 provides background on conceptual modeling and details the structured literature review. This is followed, in Section 3, by a discussion of the findings and implications extracted from the review. Section 4 discusses the implications for the continued evolution of conceptual modeling research. Section 5 concludes the paper. Appendix A (supplement) provides the entire list of papers considered. Appendix B details the results for the topic analysis by year. Appendix C identifies the top 40 bigrams and trigrams per year.

## 2 PRIOR RESEARCH AND LITERATURE REVIEW METHOD

This section reviews prior work on reviewing and analyzing conceptual modeling, as well as provides an overview of the approach taken in this paper.

### 2.1 Related work: prior reviews of conceptual modeling

Conceptual modeling can be traced to the first uses of organizational process flowcharts in the 1920s [47]. These graphical notations represented organizational decision logic, flow of resources and activities, becoming the precursors of business process models of today. Since the 1950s these diagrams increasingly contained data objects and flows of information through systems. Possibly the first mention of the need for graphical conceptual data models was made by Young and Kent in 1958 [244]: "the graphical presentation, which can be modified to suit the needs of the user (e.g., by including descriptive labels), should be helpful in determining the best organizational files and subroutines and in providing a check on redundant and superfluous information." (p.479).

Conceptual modeling, as a discipline, emerged in the 1970s in response to the emergence of new technologies, such as hierarchical, network and relational databases, as well as increased societal reliance on information systems. With the use and failures of databases and other information systems, it became clear that it was important to correctly capture real-world facts and rules, irrespective of technological solutions. In the data management area, the notion of separating logical, physical, and conceptual layers emerged [116, 209]. Through pioneering works of Abrial [2], Bachman [13-15] Codd [43], Chen [38],

Sundgren [218], Olle [168], Sibley [207], Nijssen [163-166], Kent [113], Bubenko [26, 27], and others, foundational notions of conceptual modeling and first notations and constructs were introduced. A new field of conceptual modeling was born.

Throughout its 50-year history, there have been many efforts to survey the state of the art in the area. These efforts emerged from recognizing the importance and usefulness of conceptual modeling and the various ways in which conceptual modeling has progressed. Early review efforts focused on the traditional application of conceptual modeling within the context of database design [101, 177] and process management [183]. Popular review topics included: ERDs and EERDs [4, 177, 212, 222]; UML [51]; conceptual modeling and domain ontologies [148, 180, 216, 217, 230]; and BPMN [250]. Recent research considers emergent applications of conceptual modeling within the contexts of big data [214], analytics [159, 160], machine learning [141, 247], corporate social responsibility [53], and others. These publications, although valuable, are highly focused.

In addition, significant frameworks, which seek to capture the essence of conceptual modeling and suggest future advances and research opportunities, have been developed [54, 76, 146, 158, 184, 236]. However, they have been derived mainly by considering the state of the art in conceptual modeling but lack a comprehensive survey of conceptual modeling publications from which to propose a more granular guidance for future research to the conceptual modeling community.

Prior reviews of conceptual modeling research were generally undertaken from a given disciplinary perspective. From an information systems perspective, Wand and Weber [236] assessed past conceptual modeling research and suggested focusing on evaluation, rather than development of new conceptual modeling grammars. (A *conceptual-modeling grammar* (also known as conceptual modeling language, or meta-model) is a formal specification for the creation of conceptual models. It comprises of constructs and rules that prescribe how to combine the constructs to model real-world domains [236]. Recker et al. [184] conducted a focused review of publications in information systems, which led to the formulation of a Framework for Conceptual Modeling in the Digital World. This work suggested that conceptual modeling scripts become tools of mediation between digital and physical systems. Similarly, Frank et al. [76] conducted a synoptical review of modeling publications in the journal *Business and Information Systems Engineering*. From a retrospective analysis, they suggested fruitful research opportunities, especially those dealing with emerging and new phenomena.

Some review efforts sought to combine the perspectives and analyze a wide spectrum of research in computer science, software engineering, information systems, and other relevant domains. These studies typically focused on a particular topic or modeling language. For example, Aguirre-Urreta and Marakas [4] conducted a survey of semantic data modeling techniques (specifically, extended entity-relationship vs object-oriented modeling), based on a wide range of journal articles from computer science, software engineering and information systems areas. Molina et al. [154] conducted a review of conceptual modeling of groupware systems by analyzing the software engineering and human–computer interaction literature. This survey identified several popular groupware systems notations, including the ConcurTaskTrees (CTT) [176], Group Task Analysis (GTA) Framework [228], Collaborative Usability Analysis (CUA) notation [178], and a task analysis method called the Multiple Aspect Based Task Analysis (MABTA) [130].

Other reviews of conceptual modeling are based on surveys of partitioners, as opposed to the literature alone. Notable insights on the usage of UML, for example, were made by Dobing and Parsons [58], who showed that UML is not only a language used for software engineering but also for database design. Similar analysis was conducted for other languages. For example, trends in the usage of BPMN were investigated by Compagnucci et al. [45], Rolon et al [191], Bork et al. [25], and Muehlen et al. [253].

Recognizing important historic and cultural differences in the usage of conceptual modeling, some reviews were sensitive to specific contexts and cultures. Davies at el. [51] conducted a survey of modeling with the focus on those used in Australia. Fettke [71] targeted German practitioners and also compared German and Australian modeling traditions. By combining different perspectives and sources of information, these reviews generated valuable theoretical and pedagogical insights, such as the frequency of different element usage, or the popularity of different languages.

Most reviews of conceptual modeling research focused on publications in journals, rather than conferences. This choice is understandable. Scientific journals, commonly having multiple review cycles, are generally considered to offer more rigorous, validated, and practically dependable knowledge. At the same time, when the objective is to identify emerging trends, it is also important to consider conceptual modeling conferences.

Much prior work has been based on manual extraction, and often, a manual analysis of the literature. For example, Recker et al. [184] coded the articles as exemplifying engrained assumptions in conceptual modeling (e.g., modeling conducted by professional analysts. Frank et al. [76] manually extracted modeling publications from the *Business & Information Systems Engineering* journal. A manual process is generally preferred because it permits the authors to carefully examine each publication for inclusion and relevance, as well as to classify a paper based on a given coding schema.

With the expansion of the literature, and its online availability, it becomes increasingly difficult to capture the full spectrum of work in conceptual modeling based on manual coding. Automated approaches to a literature review are increasing across all research fields [233]. For example, Harer and Fill [95] conducted a fully automated analysis of modeling publications from eight computer science and software engineering journals and the International Conference on Conceptual Modeling (ER). They used Latent Dirichlet Allocation (LDA) modeling and identified the evolution of modeling topics over time, although they did not propose a comprehensive research agenda stemming from these literature findings. In this research, we also adopt LDA as an approach to the literature review.

## 2.2 Structured Literature Review Method for Analyzing Conceptual Modeling Publications

### *2.2.1 General assumptions and approach*

To analyze conceptual modeling research, we identified the relevant literature and performed a topic analysis to extract the most frequent topics that occurred by year. An overview of this structured review process is shown in Figure 1. We manually pre-screened each paper to ensure their principal contribution is conceptual modeling, which is challenging using automated approaches alone. To add validity to the insights, we also combined these with a qualitative analysis.

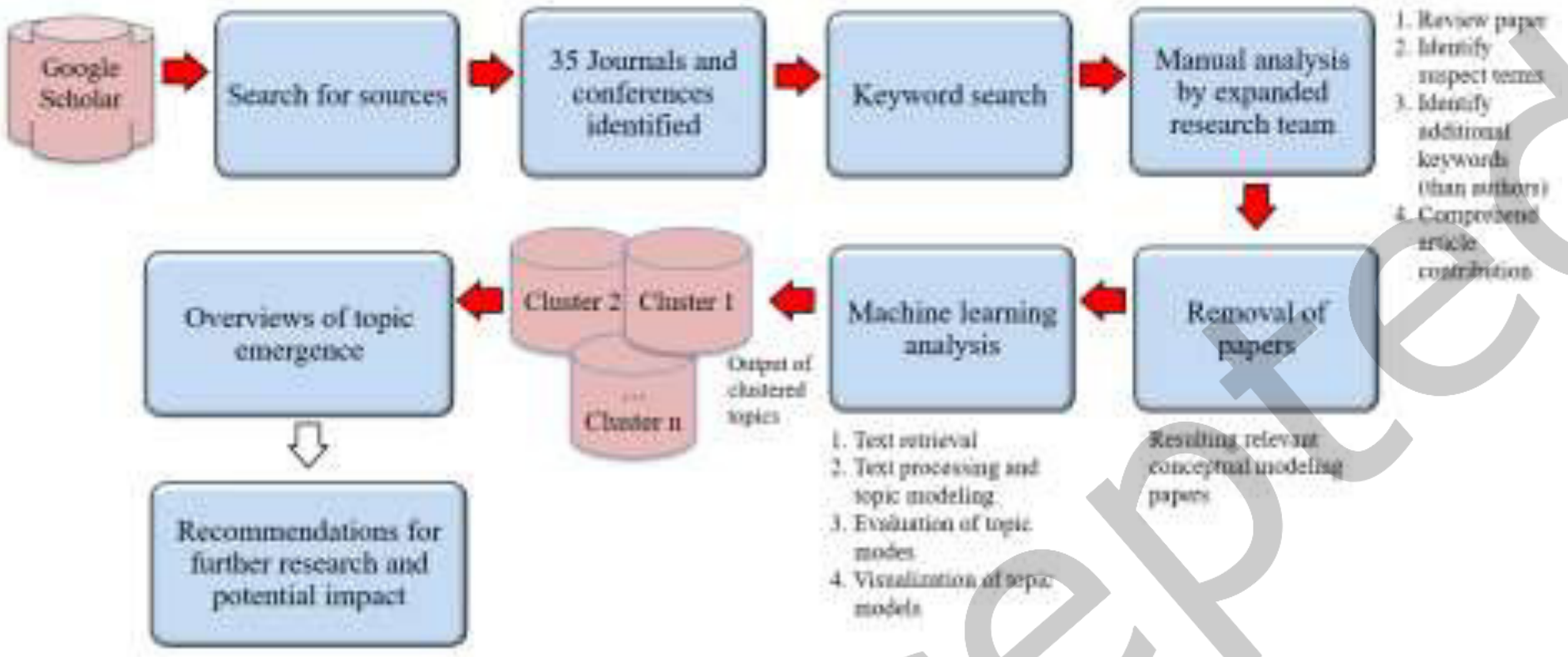

Figure 1. Structured survey of research on conceptual modeling

To appreciate and analyze the enormous amount of research that has been conducted on conceptual modeling, we performed an inclusive structured literature review. We adopted a multidisciplinary perspective, with our sources and analysis reflecting a broad spectrum of topics, themes, and trends in conceptual modeling. We also included several conferences, which offer insights into emerging research, and permit the capture of research results which, for various reasons, result in journal articles. Although our choice of literature sources is not limited to a given discipline, our disciplinary ties are mainly to the information systems area. We sought to mitigate any biases arising from adopting a particular perspective by having 6 coders (4 research assistants and 2 paper authors) and using a wide array of publication sources (35 in total).

### *2.2.2 Identification of relevant literature*

To undertake a comprehensive review, we first identified a list of sources of relevant literature. As typical of topical literature reviews (e.g., [51, 60, 124, 184, 243]), our focus was on journal publications, since academic journals typically involve multiple review cycles, leading to rigorous results. We selected a list of journals identified in prior research [4, 51, 184, 196, 247] and surveys of researchers working in the conceptual modeling area as further outlets for conceptual modeling research (e.g., https://sigsand.com/sigsand-journal-list-survey/). We also included publications from well-recognized conferences that specialize in conceptual modeling topics: the International Conference on Conceptual Modeling (ER Conference), the International Conference on Advanced Information Systems Engineering (CAiSE) and Exploring Modeling Methods for Systems Analysis and Development (EMMSAD). To capture research on behavioral and managerial implications of conceptual modeling, as well as additional empirical work on conceptual modeling languages, we added publications from two leading information systems conferences: the International Conference on Information Systems (ICIS), and Americas Conference on Information Systems (AMCIS). Although not exhaustive, these conferences still facilitated the inclusion of a wider spectrum of conceptual modeling topics.

#### *2.2.3 Keyword search for relevant topics by year*

We used keywords to search databases that contained full-text papers related to conceptual modeling in the identified journals and conferences. We used full-text search as opposed to a more common search in abstracts and keywords searching since "searching full text is more likely to find relevant articles than searching only abstracts" [132, p.1].

The terms used for the search were: "conceptual model," "conceptual modeling grammar," "ontology," and meaningful variations of these terms. Table 1 provides examples of other keywords used. We manually reviewed each search result to ensure we were: capturing an appropriate topic; identifying an appropriate paper; and interpreting the results within the entire context of the paper. We trained 4 research assistants to code the papers as candidates for inclusion. One student had completed a master's degree; two were current masters students; and one was an advanced bachelor student. A co-author supervised this effort, developing the explicit inclusion protocol shown in Table 1, to ensure process consistency, transparency, and replicability [29, 220, 233].

The result of this process included 3,910 papers published by January 2021. The literature review inclusion protocol is also summarized in Table 1. We then added two additional relevant sources: the International Conference on Advanced Information Systems Engineering (CAiSE) and the Enterprise Modelling and Information Systems Architectures Journal (EMISAJ).

Table 1. Literature review inclusion protocol for conceptual modeling (CM)

| Evidence (paper to include) | Action (taken by review team) |
|---|---|
| Term "Conceptual modeling" or any of the conceptual modeling keywords (below) are in title, abstract or keywords | Review the paper to confirm that the term deals with CM; if yes, include the paper in the analysis. |
| Conceptual model (no -ing) in the title, abstract or keywords | Review the paper to assess whether the researchers really mean CM or use CM to refer to a theory or research framework rather than for purposes of analysis and design. |
| Conceptual modeling or any of the keywords above are NOT in the title, abstract or keywords | Assess whether at least 1-2 paragraphs of the paper deal specifically with CM. |
| Additional considerations | Consider the presence of CM diagrams which illustrate CM concepts (UML, BPMN, ER); if the diagrams exist, consider the paper for inclusion and review the remainder of it. |
| Sample conceptual modeling-related terms: "Conceptual modeling" , Conceptual modelling" , "domain ontology" , "general ontology" , "ontology engineering" , "BPMN" , "UML" , Unified modeling language" , "entity relationship diagrams" , "Conceptual data modeling", "Conceptual data modelling" , "Conceptual process modeling" , "Conceptual process modelling" , "database design", "Entity-relationship model" , "Entity-relationship (ER)" , "data modeling" , "data modelling" , "object-oriented analysis" , "process-oriented analysis" , "conceptual database design" , "requirements modeling" , "requirements modelling" , "conceptual data model" , "object-role modeling" , "ER model" , "Business Process Model and Notation", "conceptual schema" | |

The final number of papers analyzed was 5,303 across 35 journals and conferences collected over the period 1976 to 2022. The sources of our analysis are summarized in Table 2.

Table 2: Sources of publications for structured literature review

| Source No. | Journals and Conferences | Number of Papers |
|---|---|---|
| 1 1 | International Conference on Conceptual Modeling | 2,062[1] |
| 2 | International Conference on Advanced Information Systems Engineering (CAiSE) | 1,213 |
| 3 | Exploring Modeling Methods for Systems Analysis and Development (EMMSAD) | 461 |
| 4 | Data & Knowledge Engineering (DKE) | 235 |
| 5 | International Journal on Software and Systems Modeling (SoSyM) | 186 |
| 6 | Enterprise Modelling and Information Systems Architectures Journal (EMISAJ) | 180 |
| 7 | Information Systems | 89 |
| 8 | IEEE Transactions on Software Engineering | 84 |
| 9 | Journal of Database Management (JDM) | 82 |
| 10 | Decision Support Systems (DSS) | 82 |
| 11 | IEEE Transactions on Knowledge and Data Engineering (TKDE) | 75 |
| 12 | Communications of the ACM (CACM) | 56 |
| 13 | Communications of the Association for Information Systems (CAIS) | 53 |
| 14 | International Conference on Information Systems (ICIS) | 48 |
| 15 | Database for Advances in Information Systems | 46 |
| 16 | ACM Transactions on Database Systems | 44 |
| 17 | Americas Conference on Information Systems (AMCIS) | 42 |
| 18 | Requirements Engineering | 33 |
| 19 | Information Sciences | 29 |
| 20 | European Journal of Information Systems (EJIS) | 28 |
| 21 | Journal of the Association for Information Systems (JAIS) | 20 |
| 22 | Information and Management | 19 |
| 23 | Journal of Management Information Systems (JMIS) | 17 |
| 24 | MIS Quarterly (MISQ) | 17 |
| 25 | Information Technology and Management | 14 |
| 26 | Information Systems Journal (ISJ) | 14 |
| 27 | Scandinavian Journal of Information Systems (SJIS) | 13 |
| 28 | Information Systems Research (ISR) | 13 |
| 29 | Journal of Information Technology (JIT) | 13 |
| 30 | Australasian Journal of information systems | 12 |
| 31 | International Journal of Information Management | 8 |
| 32 | Journal of Strategic Information Systems (JSIS) | 7 |
| 33 | ACM Journal of Data and Information Quality | 4 |
| 34 | Decision Sciences | 3 |
| 35 | Data and Information Management | 1 |
| | Total | 5,303 |

[1] Full-text papers for the International Conference on Conceptual Modeling prior to 2005 were not available at the time of retrieval.

## 2.3 Topic analysis for evolution

We analyzed the full text of the publications we retrieved (as opposed to abstracts only). Full-text articles and abstracts are structurally different [44]. Abstracts are comprised of shorter sentences and very succinct text presenting only the most important findings. Studies have shown that text mining efforts limited to abstracts lack important knowledge present in the full text documents [239].

To understand how topics evolved, we applied natural language processing (NLP) and topic analysis on a yearly basis. Using NLP has become increasingly popular to uncover useful information from large bodies of text and also allows for validation of findings through replication [1]. For most of the NLP analysis we used Latent Dirichlet Allocation (LDA), a common analysis technique that has been widely applied to natural language processing, social media analysis, and information retrieval, For example, it has been used to analyze scientific publications during the early phase of the COVID-19 pandemic [5], analyze the evolution of information systems business value research [249], assess the value of editorial reviews in user-generated content platforms [56], and investigate twitter data in real-time during a natural disaster [251]. The steps involved in the LDA analysis are described in Table 3 and illustrated using publications in the year 2000.

Table 3. Steps in topic analysis (Year 2020)

| Step | Description | Example |
|---|---|---|
| a | Create a bucket of text to analyze (here, the analysis is at the year level) | All 113 papers published in 2020 |
| b | Text process (remove stop words, lemmatization) and perform LDA (linear discriminant analysis) modeling (across various topics) | LDA model created with a varying number of topics from 8 to 14 |
| c | Evaluate topic models | Coherence score calculated for each of the models in (b) |
| d | Select optimal number of topics (i.e., highest coherence score). Topic 2 is comprised of the following top 10 terms with their respective weights: [(0.061*"process" + 0.019*"datum" + 0.019*"model" + 0.015*"activity" + 0.013*"business" + 0.011*"event" + 0.011*"log" + 0.010*"time" + 0.010*"task" + 0.010*"trace")] Each topic has a vector of terms with respective weights. | The model with highest coherence score selected. In this case, 8 topics (coherence score: 0.41) (Figure 2) |
| e | Visualize topics created and identify top terms per topic | Cluster 2 refers to process modeling/ mining |
| f | Use a language model (Doc2Vec) to find related topics. The language model was built on 2,555 full papers. | ['process', 'bpmn'] have a positive resemblance with the top 5 terms ['EPC', 'DMN', 'BPEL', 'BPEL4WS', 'Meta-model'] |

Step a. To assemble a bucket of text, we used all the papers identified in a given year to create a dataset and derive the topic model for that year.

Step b. As common in NLP, we engaged in preprocessing to prepare the data for analysis [59]. Preprocessing methods play an important role in preparing the data for insights and typically comprise the first step in the text mining process [231]. Preprocessing in our case involved creating a subset of the data for analysis and eliminating stop words and lemmatization to reduce the dimensionality of the data. We used the NLTK (Natural Language Toolkit) library [134], which includes a dictionary of common English stop words to remove (e.g., the, in, a, an) and lemmatize the corpus to reduce the dimensionality of the dataset. The lemma of a word includes its base form plus inflected forms [69, 231]. For example, the words "models", "modeled' and "modeling" have "model" as their lemma. Lemmatization groups together various inflected forms of a word into their base form (e.g., "modeling" to "model") [142]. The Spacy library supports our lemmatization task by considering nouns, adjectives, verbs, and adverbs in the documents. To augment the dataset, we used n-grams or sequences of n consecutive items. For example, bigrams (e.g., conceptual model) and trigrams (e.g., business process modeling) are included in our analysis. In Appendix C of the online supplement, we list the top 40 bigrams and trigrams created based on our corpus.

After preprocessing the data, topic modeling techniques identify relationships among the text documents. We employed the popular Latent Dirichlet Allocation (LDA) method, which has been widely applied to natural language processing, social media analysis, and information retrieval [104, 124, 233]. LDA is an unsupervised probabilistic method that assumes each document can be represented as a probabilistic distribution over latent topics [104]. Most LDA models focus on topic extraction [34] by uncovering hidden structures (semantics) from a large corpus. In LDA, topics are represented by word probabilities. The words with the highest probabilities in each topic provide a good indication of the topic. Gensim (an open source python library for topic modeling) and MALLET's LDA implementations were used to run the topic modeling on the datasets (yearly papers) [104]. MALLET is a Java-based package for statistical natural language processing, document classification, clustering, topic modeling, information extraction, and other machine learning applications to text [147].

Step c. To evaluate the topic models, we computed the coherence scores for various numbers of topics to identify the optimal model. We then used pyLDAVis Python package to visualize the information contained in a topic model [208]. A good document model should provide both coherent patterns of language and an accurate distribution of words within documents [188]. We use Gensim's coherence model to calculate topic coherence for topic [134].

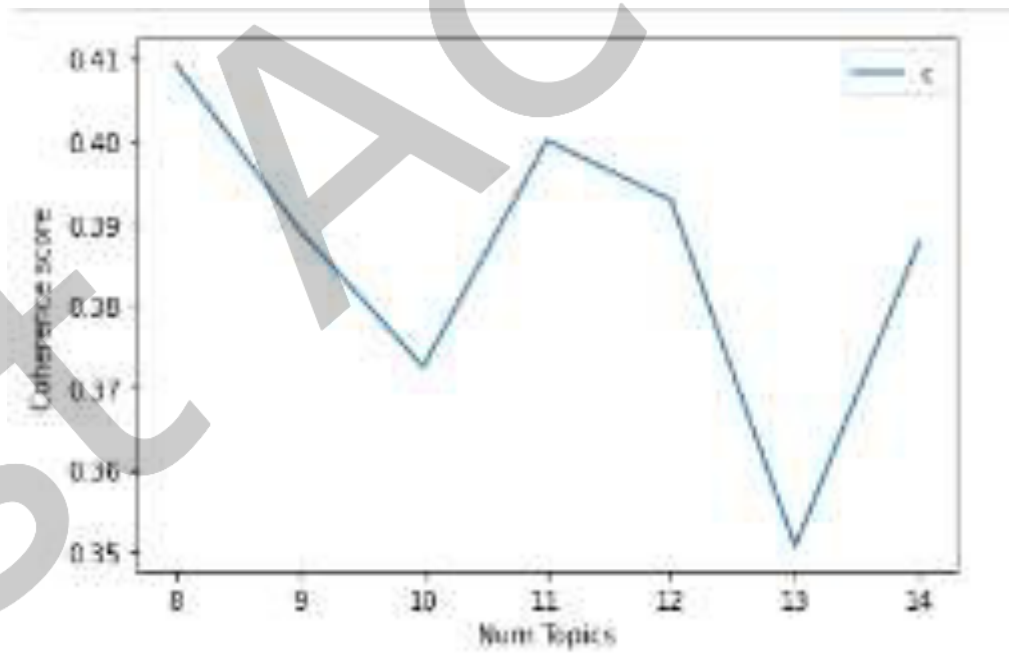

Figure 2. Coherence score by number of topics

To select the optimal number of topics, we chose the topics that gave the highest coherence score. Document models with higher topic coherence are more interpretable (i.e., words in a coherent topic have higher mutual information, thus are assumed to be related). We capped the maximum number of topics at 15 and created a visual representation, as shown in Figure 2, with an example of articles published in 2020. The image shows the coherence score at different values of N (number of topics). The same analysis was performed for each of the subsets of the corpus (i.e., content of all research papers by year). When displaying topics to users, each topic T is generally represented as a list of the M = 5, ..., N most probable words for that topic, in descending order.

Step d. To visualize the topics created and top terms per topic, we used LDAvis, a web-based interactive visualization [208]. Gensim's pyLDAVis is the most used tool to visualize the information contained in a topic model. Since each topic is embedded in a high-dimensional space, the pyLDAvis applies dimensionality reduction techniques to project each topic's high-dimensional embedding onto a 2D space [208]. Note, pyLDAvis' default principal component analysis (PCA) method, maximizes the variance of each topic's projection along the new axis using the two Principal Components, PC1 and PC2 [109]. Figure 3 is our visualization of the topic models [41, 208]. Each bubble on the left-hand side represents the marginal topic distribution. The size of the bubble, represents the marginal topic distribution (i.e., the percentage that a topic makes up in the corpus) [208]. A significant topic model has relatively large, non-overlapping bubbles scattered throughout the chart, instead of being clustered in one quadrant. A model with too many topics, typically, has many overlapping, small sized bubbles clustered in one region of the chart. On the right-hand side, the words represent the salient terms are shown. For additional illustration, we highlighted Cluster 2, showing the top 30 most salient terms that form the selected topic and their estimated term frequencies.

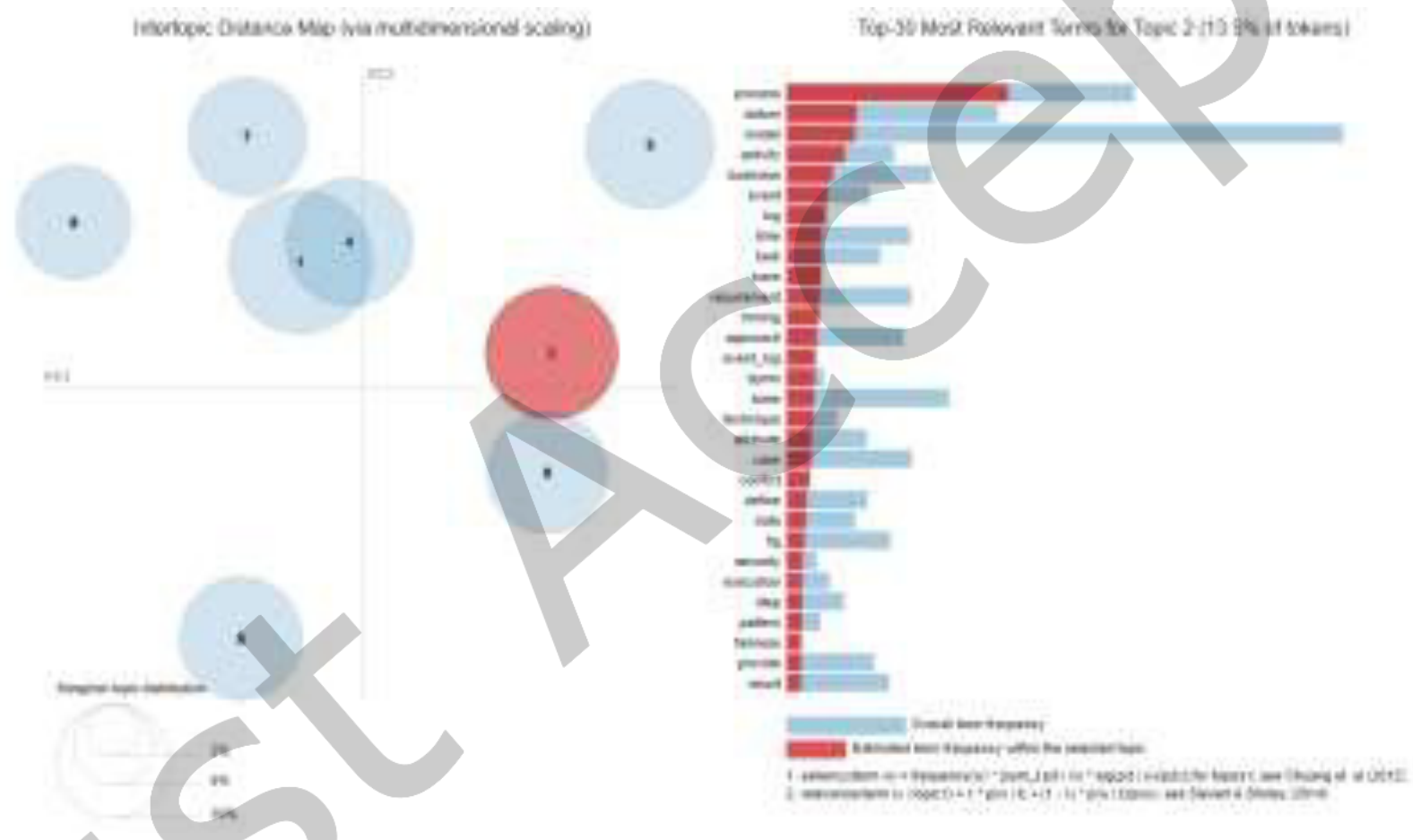

Figure 3. Visualization of Topic Models Legend: Marginal topic distribution: size of the bubble; PC1 and PC2: principle component analysis

Step f. The Doc2Vec algorithm was used to build a natural language model that created language representations from the full texts of all the conceptual modeling papers from 1975 to 2022. The model was built to complement the year-by-year analysis provided by the topic analysis. The model built identifies semantic relationships between terms [115, 125]. For example, in Figure 4, we can use terms from cluster 2 (e.g., process, bpmn, model), and use the function, *most_similar*, from the doc2vec model created, to find terms with a positive resemblance to those terms. In this case, the top 5 words are EPC (event-driven process chain), DMN (decision model and notation), BPEL (business process execution language), and meta-model, which, within the context of process models, can refer to capturing informational and behavioral aspects of business processes. Notice that although 'bpmn' and 'bpel' are syntactically different, the language model learned their semantic similarity (i.e., BPMN is used when designing and improving business processes and BPEL is the execution language to execute these business processes).

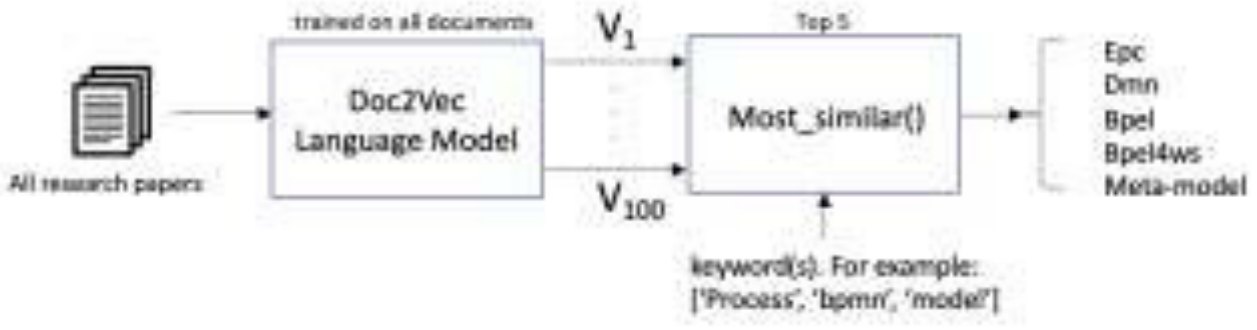

Figure 4. Language model adds context to terms in topics

With the Doc2Vec model, we built our own domain model (language model) of conceptual modeling. Based on this language model we can now estimate how close words are to each other (co-occurrence), and find similarities or synonyms, without explicitly identifying them. This model could also be used for any future analysis. We then applied our language model to the buckets of research papers by year.

## 3 RESULTS AND FINDINGS

To present the findings from our literature review analysis, we employ a mixed-method approach, where we combine insights drawn from the statistical techniques described above with manual and qualitative insights. In the manual and qualitative review, we draw upon known publications from 1970s to 2022. Our statistical analysis focuses mainly on research from 2005 to 2020. There was a three-fold increase in publication rates from 2005 onwards. Note, this does not necessarily represent an increase in the volume of research papers, although there is reported evidence of such an increase [184, 196]. This could also be a function of increased inclusion of journal publications in the databases used to extract the papers. Hence, our analysis focuses on 2005-2020 to ensure that our results are less sensitive to the varied digitization practices of academic content. The focused analysis of more recent publications further permits stronger inferences regarding emerging topics in conceptual modeling, providing practical utility to researchers wishing to identify existing gaps. At the same time, we reviewed all available papers to uncover themes and draw general conclusions regarding the state of conceptual modeling research and supplement and interpret our natural language processing findings with this manual effort. Table 4 reports the number of topics by year and the number of articles with full text found in our sample for that year.

Table 4: Topics and number of articles per year[2]

| Year | No. of articles considered | No. of articles with full text | No. of Topics |
|---|---|---|---|
| 1976-2004 | 801 | 639 | 6 |
| 2005 | 215 | 169 | 9 |
| 2006 | 207 | 171 | 10 |
| 2007 | 216 | 165 | 11 |
| 2008 | 217 | 163 | 11 |
| 2009 | 205 | 158 | 8 |
| 2010 | 220 | 157 | 10 |
| 2011 | 257 | 187 | 9 |
| 2012 | 231 | 175 | 10 |
| 2013 | 255 | 152 | 9 |
| 2014 | 207 | 139 | 10 |
| 2015 | 240 | 155 | 10 |
| 2016 | 174 | 112 | 10 |
| 2017 | 174 | 124 | 12 |
| 2018 | 228 | 132 | 8 |
| 2019 | 202 | 139 | 7 |
| 2020 | 173 | 113 | 8 |
| 2021[3] | 72 | 72 | 6 |
| 2022[4] | 51 | 51 | 8 |
| **Total** | **4,345** | **3,173** | **172** |

## 3.1 Topics for 1976 to 2004

Prior to 2005, there were 639 papers with full text extracted from the databases based on the process we followed. We followed the steps described in the methodology to summarize the significant topics that emerged. Below are the main topics (see Appendix A for the list of all papers and Appendix B for analysis of the entire literature by year).

**Topic 0 (Analysis):** terms of system, model, information, process, design, analysis, requirement, development, user, datum. This includes computer-aided software engineering [37] and improving the quality of data models [156]. A methodology to derive requirements for information systems development [198], involves cognitive fit in requirements modeling [3], or understanding and representing user viewpoints during requirements [49]. The diffusion of information systems development methods was explored [23] as well as Object-Oriented Systems Development [108]. Challenges of strategic data planning [205] were investigated as were efforts to describe organizations as sets of business processes and to derive a conceptual framework for understanding such business processes and business process modeling [110], and business process redesign [221, 238].

**Topic 1 (State flow):** terms of model, state, event, time, system, specification, object, temporal. A structured operational semantics for UML-state charts was investigated [232], as was a formalization of

[2] For some of the articles only abstracts were available, hence the overall total number of full texts.
[3] Articles included from CAiSE and EMISAJ only.
[4] Articles included from CAiSE and EMISAJ until October 2022.

UML state machines using temporal logic [193], and a method for describing the syntax and semantics of UML statecharts [107]. Tool support for verifying UML activity diagrams [65] and automatically detecting and visualizing errors in UML Diagrams [32] was developed. Additional research focused on formal semantics of static and temporal state-oriented OCL constraints [74].

**Topic 2 (Object-oriented modeling):** terms of object, class, type, model, instance, attribute, property, set, define. A multi-level view model for secure object-oriented databases were developed [16], as was an analysis of the notion and issues related to object-oriented query languages [22]. A template for defining enterprise modeling constructs was developed [169], as was formal semantics of an Entity-Relationship-Based query language [98].

**Topic 3 (Entity-relationship modeling):** terms of entity, relationship, attribute, set, database, relation, type, constraint, model, key. Papers include analyzing the entity-relationship approach to database design [161]; proposing a normal form for relational databases based on domains and keys [67], proposing an algebra for a general entity-relationship model [172], representing extended entity-relationship structures in relational databases [144], or discussing how to map an entity-relationship schema into a SQL schema [123]. Justification for the inclusion of dependency normal form was investigated [128]. Researchers also investigated computational problems related to the design of normal form relational schemas [17] and the design of relational database schemata, generally [248].

**Topic 4 (Knowledge modeling):** terms of type, knowledge, model, conceptual, concept, information, system, ontology, language, domain. Research on domain ontologies included a declarative approach for reusing them [152] and their grammatical specification [7]. Other work focused on an algebraic approach to modular construction of logic knowledge bases [204]; subtyping and polymorphism in object-role modeling [93]; and expressiveness in conceptual modeling [223].

**Topic 5 (Data modeling):** terms of database, datum, design, query, system, data, number, base, user, table. Papers include data base research [153], physical design for relational databases [73] and the sensitivity of physical design to changes in underlying factors [170]. Other efforts focused on a preliminary system for the design of DBTG data structures [83], database design principles for placement of delay-sensitive data on disks, and a case study of database design using the dataid approach [52].

Table 5. Emergence of topics from papers related to conceptual modeling

| 2005 | 2006 | 2007 | 2008 | 2009 | 2010 | 2011 | 2012 | 2013 | 2014 | 2015 | 2016 | 2017 | 2018 | 2019 | 2020 |
|---|---|---|---|---|---|---|---|---|---|---|---|---|---|---|---|
| Ontology<br>Domain<br>Semantic<br>Language | Model<br>System<br>Task<br>Knowledge | Query<br>Database<br>Tree<br>View | Model<br>System<br>Domain<br>Approach | Relationship<br>Relation<br>Object<br>Role | Datum<br>State<br>Model<br>Property | Process<br>System<br>Information<br>Method | Constraint<br>Schema<br>Database<br>Temporal | Model<br>Tool<br>Language<br>Approach | Process<br>Model<br>Business<br>Activity | Model<br>Graph<br>Rule<br>Edge | Datum<br>Information<br>Business<br>Knowledge | Pattern<br>Model<br>Type<br>Object | Model<br>Conceptual<br>Requirement<br>Quality | Query<br>Graph<br>Schema<br>Semantic | Model<br>business<br>Service<br>platform |
| Model<br>Agent<br>UML<br>System | Ontology<br>Domain<br>Owl<br>Semantic | Web<br>Domain<br>Ontology<br>Cluster | Process<br>Model<br>Business<br>Activity | State<br>UML<br>Diagram<br>Operation | Constraint<br>Class<br>Rule<br>OCL | Set<br>Ontology<br>Relation<br>Concept | Time<br>Datum<br>System<br>Property | Goal<br>Model<br>Business<br>Requirement | Datum<br>Process<br>Risk<br>Quality | Model<br>System<br>Modeling<br>Process | Model<br>Constraint<br>Class<br>OCL | Process<br>Decision<br>Model<br>Task | Result<br>Method<br>Approach<br>Technique | Model<br>State<br>Execution<br>Transition | Process<br>Model<br>Business<br>Activity |
| Constraint<br>State<br>Event<br>Specification | Node<br>Edge<br>Algorithm<br>Graph | Model<br>Datum<br>Context<br>Level | Requirement<br>Case<br>Goal<br>Method | Concept<br>Ontology<br>Web<br>Domain | Process<br>Business<br>Activity<br>Model | UML<br>Model<br>Event<br>Requirement | Query<br>User<br>Concept<br>Dimension | Datum<br>Query<br>Database<br>Time | Information<br>System<br>Model<br>Ontology | Pattern<br>Class<br>Constraint<br>UML | State<br>Node<br>Process<br>Semantic | Data<br>Conceptual<br>Approach<br>Modeling | Database<br>Query<br>Schema<br>Graph | Modeling<br>System<br>Ontology<br>Domain | Process<br>Case<br>Object<br>Set |
| Model<br>Class<br>Datum<br>Dimension | Conceptual<br>Modeling<br>System<br>Ontology | Goal<br>Requirement<br>Modeling<br>Security | Object<br>Property<br>Concept<br>Event | Ontology<br>Domain<br>OWL<br>Knowledge | Service<br>Query<br>Node<br>Web | Conceptual<br>Modeling<br>Grammar<br>Domain | Model<br>Modeling<br>Process<br>Design | Process<br>Activity<br>Model<br>Business | Model<br>Result<br>Method<br>UML | Object<br>Datum<br>State<br>Message | Ontology<br>Requirement<br>Model<br>Domain | Model<br>Component<br>Contract<br>Block | Role<br>Element<br>Class<br>UML | Datum<br>user<br>Data<br>Approach | Model<br>Language<br>Specification<br>Tool |
| Schema<br>Type<br>Relationship<br>Query | Process<br>Goal<br>System<br>Requirements | System<br>Method<br>Information<br>Design | Ontology<br>Class<br>Semantic<br>Quality | Model<br>Goal<br>Process<br>Requirement | Ontology<br>Goal<br>Relation<br>Context | Service<br>Feature<br>Web<br>Workflow | Domain<br>Ontology<br>Semantic<br>Knowledge | Ontology<br>User<br>Relation<br>Domain | Goal<br>Model<br>Requirement<br>Business | Ontology<br>Domain<br>Type<br>Conceptual | Datum<br>Type<br>Entity<br>Relationship | Graph<br>Set<br>Query<br>Node | Process<br>Model<br>Activity<br>Event | Model<br>Class<br>Language<br>UML | Model<br>Modeling<br>Task<br>Question |
| Process<br>Task<br>Goal<br>Activity<br>Model | Class<br>Object<br>Diagram<br>UML<br>Constraint | Attribute<br>Element<br>Schema<br>Fuzzy<br>Class | Datum<br>Schema<br>Dimension<br>Data<br>Warehouse | Set<br>Constraint<br>Type<br>Role<br>Property | UML<br>System<br>Analysis<br>Object<br>Message | Requirement<br>Security<br>System<br>Goal<br>Specification | Object<br>Type<br>Issue<br>Set<br>Entity | Model<br>System<br>Information<br>Modeling<br>Work | Query<br>Time<br>Object<br>Schema<br>Database | Datum<br>Schema<br>Database<br>Attribute<br>Entity | Model<br>Diagram<br>Consistency<br>UML<br>Software | Process<br>Model<br>BPMN<br>Activity<br>Business | Type<br>Level<br>Instance<br>Property<br>Model | Constraint<br>Class<br>OCL<br>Case<br>UML | Datum<br>Query<br>Model<br>Time<br>Schema |
| Spatial<br>Graph<br>Node<br>Relationship | Service<br>Process<br>Business<br>Web service | Unit<br>State<br>Pattern<br>Diagram | User<br>Service<br>Web<br>Application | Model<br>Class<br>Instance<br>Language | Model<br>Modeling<br>Language<br>Architecture | Activity<br>Diagram<br>Node<br>Scenario | System<br>Information<br>Requirement<br>Goal | Model<br>Event<br>Aspect<br>Message | Service<br>Model<br>Process<br>Node | Model<br>Result<br>Notation<br>Experiment | Process<br>Model<br>Event<br>Activity | Model<br>Process<br>Experiment<br>Comprehension | Model<br>UML<br>Type<br>Profile | Process<br>Activity<br>Model<br>BPMN | Ontology<br>System<br>Concept<br>Information<br>Domain |
| System<br>Model<br>Method<br>Construct | Schema<br>Database<br>Temporal<br>Semantic | Conceptual<br>Ontology<br>System<br>Language | Information<br>System<br>Process<br>Quality | Modeling<br>Model<br>System<br>Conceptual | Datum<br>Attribute<br>Database<br>Schema | Datum<br>Model<br>View<br>Schema | UML<br>Model<br>State<br>Modeling | Relationship<br>Schema<br>Graph<br>Node | Domain<br>Ontology<br>Class<br>Knowledge | Model<br>Case<br>Language<br>System | Model<br>Concept<br>View<br>Architecture | Model<br>Datum<br>Service<br>Cloud | Time<br>State<br>System<br>Requirement | | Model<br>System<br>Requirement<br>Simulation |
| System<br>Diagram<br>Design<br>Method | Datum<br>Type<br>Entity<br>Design | Object<br>Association<br>Property<br>Class | Database<br>Relation set<br>Schema<br>Constraint | | Information<br>Process<br>Knowledge<br>Modeling | Model<br>Class<br>UML<br>Interaction | Process<br>Model<br>Business<br>Service | Model<br>Class<br>System<br>Property | Model<br>Constraint<br>Pattern<br>Set | Goal<br>Model<br>Requirement<br>Security | System<br>Design<br>User<br>Goal | Model<br>State<br>Algorithm<br>Mining | | | |
| | | Process<br>Model<br>Business<br>Activity | Time<br>Constraint<br>Spatial<br>Semantic | | | | Model<br>Component<br>Operation<br>Simulation | | | | | System<br>Design<br>Method<br>Theory | | | |

As can be seen from the automated extraction of topics, the main theme in conceptual modeling over the years, until 2005, focused on analysis, and various types of conceptual modeling. Of note, Chen's [38] entity-relationship model was the main conceptual model used to represent an application domain. Smith and Smith [210] identified abstractions as an important way to capture semantics. Other work on data abstractions and semantic relationships were based on this work (e.g., [84, 85, 129, 138, 146, 173, 213, 240]). Efforts were underway to recognize the need to separate database design phases and to create formal ways to transform a conceptual to a logical design with the relational model becoming the standard (e.g. [222]). The emergence of object-oriented methodologies, systems development, UML, domain ontologies, grammars, and other methods began, consistent with our analysis. We can also conclude that largely, conceptual modeling was focused on modeling data and processes.

### 3.2 Topics for 2005-2020

Table 5 provides an overview of the topics that emerged from 2005-2020. Each cell provides a topic and the most representative terms for that topic. The circled terms show, for example, how the topic of ontology progressed over the 15-year analysis. In 2005, ontology appeared in connection with semantic languages. Five years later, in 2010, it was not prominent; nor was it in 2017 and 2018. In other years there is evidence of its use with respect to the web, topics related to knowledge management and, more recently, for requirements and domain concepts.

In Table 5, the main terms (by cluster) are provided; the less-frequent ones are not shown for clarity of presentation. In 2010, for example, the clustered topics can similarly be labeled as: Topic 1 (data); Topic 2 (constraints); Topic 3 (business processes); Topic 4 (web services); Topic 5 (context); Topic 6 (UML); Topic 7 (modeling); Topic 8 (schema); Topic 9 (knowledge management). The topic analysis for 2020 is presented below. For all the other years, the details of the topic analysis are given in Appendix B.

#### *3.2.1 Topic Analysis for Year 2020*

To illustrate the results of our analysis, we highlight the year 2020. The topics are described below and are indicative of the types of research carried out in 2020.

**Topic 1 (business process design and execution)** refers to model, business, service, platform, and also method, design, capability, goal, and process. Sixteen papers fall under this topic. Business Process Management Suites (BPMS) are being adopted in organizations to increase business process agility. Yet, organizations struggle to achieve agile business processes. BPMS is useful for practitioners wanting to adopt a business process management suite that addresses the difficulty of integrating with other applications [118]. Organizations operate within dynamic environments to which they need to adapt. The age of digitization requires rapid design and re-design of enterprises. The design and engineering methodology for organizations (DEMO) is an established modeling method for representing the organization domain of an enterprise. Gray et al. [86] addresses stakeholder heterogeneity by enabling transformation of a DEMO organization construction diagram (OCD) into a BPMN collaboration diagram. Currently, enterprise modeling and capability modeling facilitate the design and analysis of capabilities. Koutsopoulos et al. [119] introduces a capability change meta-model that serves as the basis for capability change. Recently, Industry 4.0 has attracted much research and development over the last decade. At its core is the need to connect physical devices with their digital representations (i.e., digital twin). Sandkuhl & Stirna [201] analyzes the suitability of enterprise modeling and capability management for the purpose of developing and management of business-driven digital twins.

**Topic 2 (business modeling and mining)** is characterized by the terms process, datum, model, activity, business, event, log, time, task, and trace. Eighteen papers fall under this topic. Ramadan et al. [182] propose a BPMN-based framework that supports the design of business processes considering security, data-minimization and fairness requirements. Camargo et al. [31] present an accuracy-optimized method to discover business process simulation models from execution logs. Several process mining techniques discover models for predictive analyses. These techniques need an appropriate time step size, the selection of which, thus far, has been an ad-hoc and manual endeavor. Pourbafrani et al. [179] propose a novel semi-automated time-granularity detection framework and highlight the importance of using accurate granularity in time step selection. Process mining aims to obtain insights from event logs to improve business processes. In complex environments with large variances in process behavior, analyzing and making sense of complex processes becomes challenging. Insights into such processes can be obtained by identifying sub-groups of traces (cohorts) and studying their differences. Leemans et al. [126] introduces a framework that considers ordering of activities in traces (control flow), the relative frequency of traces (stochastic perspective), and cost.

**Topic 3 (process analysis)** is characterized by the terms process, case, object, set, instance, state, event, model, context, and datum. Fourteen papers fell under this topic. Andrews et al. [9] presents concepts for enabling context switching at runtime for object-aware process management and discusses use cases in which context switching capabilities can be utilized. Rodrigues et al. [189] explores the view of occurrents as transitions between situations and propose a framework for the ontological analysis of occurrents. Andree et al. [8] developed an exception handling technique for fragment-based case management (fCM) for handling unknown events.

**Topic 4 (modeling languages)** is characterized by the terms model, language, specification, tool, class, element, type notation, object, and modeling. Ten papers fall under this topic. Bork et al. [25] provide a Systematic Literature Review aimed to analyze published standard modeling language specifications such as Business Process Model and Notation and the Unified Modeling Language. This survey provides a foundation for research aiming to increase consistency and improve comprehensiveness of information systems modeling languages. The survey reveals heterogeneity in: (i) the modeling language concepts being specified; and (ii) the techniques being employed for the specification of these concepts. Zolotas [252] bridges a proprietary UML modeling tool used for model-based development of safety-critical systems with an open-source family of languages for automated model management.

**Topic 5 (empirical evaluations)** is characterized by the terms model, modeling, task, question, group, subject, conceptual, result, study, and experiment. Nine papers fall under this topic. Bernardez [21] evaluates whether Software Engineering students enhance their conceptual modeling performance after several weeks of practicing mindfulness. Verdonck et al. [230] study the extent to which the pragmatic quality of ontology-driven models is influenced by the choice of a particular ontology, given a certain understanding of that ontology.

**Topic 6 (data models)** is characterized by the terms datum, query, model, data, time, set, schema, database, and table. Thirteen papers fall under this topic. Wang et al. [237] propose a Deep Temporal Multi-Graph Convolutional Network (DT-MGCN) model that integrates graph generation component with spatial-temporal component to capture the dependencies between crime and various external factors. Hartmann et al. [96] implements an efficient approach for dynamic alternative route planning that can respond to road network changes. For data models used in big data analysis, such as Multilayer Networks, there is a need to transform the user/application requirements using a modeling approach such as the extended entity relationship (EER). Komar et al. [117] show how the EER approach can be leveraged for modeling given data to generate Multi-layer Networks (MLN)s and appropriate analysis expressions on them.

**Topic 7 (ontology)** is characterized by the terms ontology, system, concept, information, conceptual, domain, vulnerability, knowledge base, and research. Fourteen papers fall under this topic. For example, Syed [219] present the Cybersecurity Vulnerability Ontology (CVO), a conceptual model for formal knowledge representation of the vulnerability management domain. Lukyanenko et al. [139] propose a General Systemist Ontology (GSO) as a foundation for developing information technologies where the application could benefit from a systems perspective.

**Topic 8 (software engineering)** is characterized by the terms model, system, requirement, simulation, base, design, software, and engineering. Seven papers fall under this topic and provide a systematic mapping and solid basis for classifying approaches to systems modeling language (SysML) [241]; Artifact-Based Workflows for Supporting Simulation Studies [195]; and a Toolbox for the Internet of Things-Easing the Setup of IoT Applications [77]. Within the context of digital transformation, speeding up the time-to-market of high-quality software products is a big challenge. Software quality correlates with the success of requirements engineering sessions (e.g., software analysts collect relevant material). Comprehensible requirements need to be specified for software implementation and testing. Many of these activities are performed manually, causing process delays and software quality issues, such as reliability, usability, comprehensibility. Ruiz & Hasselman [194] propose a framework for automating the tasks of requirements specification.

### *3.2.2 Prevalence of terms*

In addition to the analysis by topic and by year, we also analyzed the prevalence of terms across the years. Table 6 summarizes the terms, capturing the conceptual modeling interest, as it progressed over the last 15 years. Clearly, some are reoccurring, others fad over time, and yet others, evolve. As expected, terms related to models, systems, and processes were consistent over the years of study. The terms information, goal, class, ontology, domain, database, and schema were likewise found over the years.

Table 6. Terms by year

| Term | 2005 | 2006 | 2007 | 2008 | 2009 | 2010 | 2011 | 2012 | 2013 | 2014 | 2015 | 2016 | 2017 | 2018 | 2019 | 2020 |
|---|---|---|---|---|---|---|---|---|---|---|---|---|---|---|---|---|
| Activity | X | | X | X | | X | X | | X | X | | | X | | X | X |
| Agent | X | | | | | | | | | | | | | | | |
| Algorithm | | X | | | | | | | | | | | | | | |
| Analysis | | | | | | X | | | | | | | | | | |
| Application | | | | X | | | | | | | | | | | | |
| Approach | | | | X | | | | | | | | | X | | | |
| Architecture | | | | | | X | | | | | | | | | | |
| Aspect | | | | | | | | | X | | | | | | | |
| Association | | | X | | | | | | | | | | | | | |
| Attribute | | | X | | | | | | | | X | | | | | |
| Block | | | | | | | | | | | | | X | | | |
| BPMN | | | | | | | | | | | | | X | | X | |
| Business | | X | X | X | | X | | | X | X | | X | X | | | X |
| Case | | | | | | | | | | | | | | | | X |
| Class | X | X | X | X | X | | X | | | X | X | X | | X | X | |
| Cloud | | | | | | | | | | | | | X | | | |
| Cluster | | | X | | | | | | | | | | | | | |
| Comprehension | | | | | | | | | | | | | X | | | |
| Concept | | | | | X | | X | X | | | | | | | | X |
| Conceptual | | X | X | | X | | X | | | | | | X | X | | |
| Consistency | | | | | | | | | | | | X | | | | |
| Constraint | X | X | | X | X | X | | X | | X | X | X | | | X | |
| Construct | X | | | | | | | | | | | | | | | |
| Context | | | X | | | | | | | | | | | | | |
| Contract | | | | | | | | | | | | | X | | | |
| Data | | | | X | | | | | | | | | X | | | |
| Database | | X | X | X | | X | | X | X | X | X | | | X | | |

| Term | 2005 | 2006 | 2007 | 2008 | 2009 | 2010 | 2011 | 2012 | 2013 | 2014 | 2015 | 2016 | 2017 | 2018 | 2019 | 2020 |
|---|---|---|---|---|---|---|---|---|---|---|---|---|---|---|---|---|
| Datum | X | X |  | X |  | X | X | X | X | X | X | X |  |  | X |  |
| Decision |  |  |  |  |  |  |  |  |  |  |  |  | X |  |  |  |
| Design | X | X | X |  |  |  |  | X |  |  |  | X |  |  |  |  |
| Diagram | X | X | X |  |  |  | X |  |  |  |  | X |  |  |  |  |
| Dimension | X |  |  | X |  |  |  | X |  |  |  |  |  |  |  |  |
| Domain | X | X | X | X | X |  | X | X | X | X | X | X |  |  |  | X |
| Edge |  | X |  |  |  |  |  |  |  |  | X |  |  |  |  |  |
| Element |  |  | X |  |  |  |  |  |  |  |  |  |  | X |  |  |
| Entity |  | X |  |  |  |  |  | X |  |  | X | X |  |  |  |  |
| Event | X |  |  | X |  |  | X |  |  |  |  | X |  | X |  |  |
| Execution |  |  |  |  |  |  |  |  |  |  |  |  |  |  | X |  |
| Experiment |  |  |  |  |  |  |  |  |  |  | X |  | X |  |  |  |
| Fuzzy |  |  | X |  |  |  |  |  |  |  |  |  |  |  |  |  |
| Goal | X |  | X |  | X | X | X |  | X | X | X | X |  |  |  |  |
| Grammar |  |  |  |  |  |  | X |  |  |  |  |  |  |  |  |  |
| Graph | X | X |  |  |  |  |  |  | X |  | X |  | X | X | X |  |
| Information |  |  | X | X | X | X | X | X | X | X |  | X |  |  |  | X |
| Instance |  |  |  |  | X |  |  |  |  |  |  |  |  | X |  |  |
| Interaction |  |  |  |  |  |  | X |  |  |  |  |  |  |  |  |  |
| Issue |  |  |  |  |  |  |  | X |  |  |  |  |  |  |  |  |
| Knowledge |  | X |  |  |  | X |  | X |  | X |  | X |  |  |  |  |
| Language | X |  |  |  | X | X |  |  | X |  | X |  |  |  |  | X |
| Level |  |  | X |  |  |  |  |  |  |  |  |  |  |  |  |  |
| Message |  |  |  |  |  | X |  |  |  |  |  |  |  |  |  |  |
| Method | X |  | X |  |  |  | X |  |  |  |  |  |  |  |  |  |
| Mining |  |  |  |  |  |  |  |  |  |  |  |  | X |  |  |  |
| Model | X | X | X | X | X | X | X | X | X | X | X | X | X | X | X | X |
| Modeling |  | X |  |  |  | X | X | X |  |  | X |  |  |  |  | X |
| Node | X | X |  | X |  |  |  |  |  |  |  | X | X |  |  |  |
| Notation |  |  |  |  |  |  |  |  |  |  | X |  |  |  |  |  |
| Object |  | X | X |  |  | X |  |  |  |  |  |  | X |  |  |  |
| OCL |  |  |  |  |  | X |  |  |  |  |  | X |  |  | X |  |
| Ontology | X | X | X | X | X | X | X | X | X | X | X | X |  |  | X | X |
| Operation |  |  |  |  | X |  |  | X |  |  |  |  |  |  |  |  |
| Owl |  | X |  |  | X |  |  |  |  |  |  |  |  |  |  |  |
| Pattern |  |  | X |  |  |  |  |  |  |  | X |  | X |  |  |  |
| Process | X | X | X | X | X | X | X | X | X | X | X | X | X | X | X | X |
| Profile |  |  |  |  |  |  |  |  |  |  |  |  |  | X |  |  |
| Property |  |  | X | X |  |  |  | X |  |  |  |  |  |  |  |  |
| Quality |  |  |  | X |  |  |  |  |  | X |  |  |  | X |  |  |
| Query | X |  | X |  |  |  |  | X | X | X |  |  | X | X | X |  |
| Question |  |  |  |  |  |  |  |  |  |  |  |  |  |  |  | X |
| Relation |  |  |  |  | X | X | X |  | X |  |  |  |  |  |  |  |
| Relation Set |  |  |  | X |  |  |  |  |  |  |  |  |  |  |  |  |
| Relationship | X |  |  |  | X |  |  |  | X |  |  |  |  |  |  |  |
| Requirements |  | X | X | X | X |  | X | X | X | X | X | X |  | X |  | X |
| Risk |  |  |  |  |  |  |  |  |  | X |  |  |  |  |  |  |
| Role |  |  |  |  | X |  |  |  |  |  |  |  |  |  |  |  |
| Rule |  |  |  |  | X |  |  |  |  |  |  |  |  |  |  |  |
| Schema | X | X | X | X | X | X |  | X | X | X | X |  |  |  |  |  |
| Security |  |  | X |  |  |  | X |  |  |  | X |  |  |  |  |  |

| Term | 2005 | 2006 | 2007 | 2008 | 2009 | 2010 | 2011 | 2012 | 2013 | 2014 | 2015 | 2016 | 2017 | 2018 | 2019 | 2020 |
|---|---|---|---|---|---|---|---|---|---|---|---|---|---|---|---|---|
| Semantic | X | X | | X | | | | X | | | | X | | | X | |
| Service | | X | | | | X | X | | | | | | X | | | |
| Set | | | | | X | | X | | | X | | | X | | | X |
| Simulation | | | | | | | | X | | | | | | | | |
| Software | | | | | | | | | | | | X | | | | |
| Spatial | X | | | | | | | | | | | | | | | |
| Specification | X | | | | | | X | | | | | | | | | |
| State | X | | | | | X | | | | | X | X | X | X | X | |
| System | X | X | X | X | X | X | X | X | X | X | X | X | X | X | X | X |
| Task | X | X | | | | | | | | | | | X | | | X |
| Technique | | | | | | | | | | | | | | X | | |
| Temporal | | X | | | | | | X | | | | | | | | |
| Theory | | | | | | | | | | | | | X | | | |
| Time | | | | X | | | | X | | X | | | | X | | |
| Transition | | | | | | | | | | | | | | | X | |
| Tree | | | X | | | | | | | | | | | | | |
| Type | X | X | | | | | | X | | | X | X | X | X | | |
| UML | X | X | X | | X | X | X | X | | | X | X | | X | X | |
| User | | | | X | | | | | X | | | | | | X | |
| View | | | X | | | | | | | | | X | | | | |
| Warehouse | | | | X | | | | | | | | | | | | |
| Web | | | X | X | X | X | X | | | | | | | | | |
| Web Service | | X | | | | | | | | | | | | | | |
| Work | | | | | | | | | X | | | | | | | |

Some terms have been consistently popular. These include such general terms as data, information, database, requirements, and process. Some terms that appeared in multiple years were sometimes confined to a small number of years. For example, the term, relation, had consistency between 2008 to 2013, as relational databases continued to mature, and become standardized. (Thereafter, NoSQL and NewSQL became popular). Other topics have a sporadic influence, such as knowledge, concept, or property. Still, others emerged as relevant but only for one year, such as message, interaction, and element. Some of these were more notable in early years but may have been absorbed in different ways. For example, construct and context are known to be important concepts in conceptual modeling and have been prominent in many research endeavors throughout the years.

### 3.3 Additional analysis

The above analysis is based on the amalgamation of conference and journal papers on conceptual modeling. Collectively, it presents the overall themes, topics, and trends, common across the entire conceptual modeling community. However, conceptual modeling community is highly diverse and heterogeneous. It draws upon the concepts, theories, challenges and problems from computer science, software engineering, management information systems, design science, as well as scientific domains (e.g., genomics), being influenced by such disciplines as philosophy, cognitive science, psychology, linguistics, semiotics, among others.

Very few attempts have been made to take a full stock of the diversity of conceptual modeling, the different perspectives that comprise it, or the identification of the common thread among these perspectives. While doing so directly is beyond the scope of this paper, we contribute to these goals by conducting a focused analysis of two specific outlets for conceptual modeling: the *International Conference on Advanced Information Systems Engineering (CAiSE)* and the open access journal, *Enterprise Modelling and Information Systems Architectures (EMISAJ)*.[5]

The *Enterprise Modelling and Information Systems Architectures* is a journal that, although accepting all conceptual modeling research, has a historic focus on Enterprise Modelling, Information Systems Architectures and Business Modeling (Modellierung betrieblicher Informationssysteme). It originated in a prolific German modeling community. The community has been known for its contributions to business information systems design and interest in organizational information systems.

The *International Conference on Advanced Information Systems Engineering*, on the other hand, is a premier conference with a traditional focus on "Information Systems Engineering with a special emphasis on the theme of Cyber-Human Systems" (emphasis in the original, see: caise23.svit.usj.es). The first Conference on "Advanced Systems Engineering", CASE'89, was arranged during in May 1989, jointly by SISU in Stockholm, Sweden and has been active ever since. The conference was originally organized by prominent Nordic conceptual modeling community lead by the Swedish Institute for Systems Development (SISU) in co-operation with the Swedish Society for Information Processing SSI [190]. Attesting to the importance of understanding the distinct voices within community, Rolland et al. [190] explained that none of the existing venues was providing a broad integration of modeling with issues of information systems development and its pertinent social factors. The isolation of these two sources permitted a comparison of the general corpus with the sources that have a disciplinary focus – computer science and software engineering. Furthermore, by conducting a more focused analysis, we were able to surface some of the niche topics, which were overshadowed by the mainstream topics in the general corpus. Table 7 summarizes the topic analysis.

[5] These specific outlets were suggested by the very helpful reviewers as additional sources to analyze.

Table 7. Topics from papers from CAiSE and EMISAJ

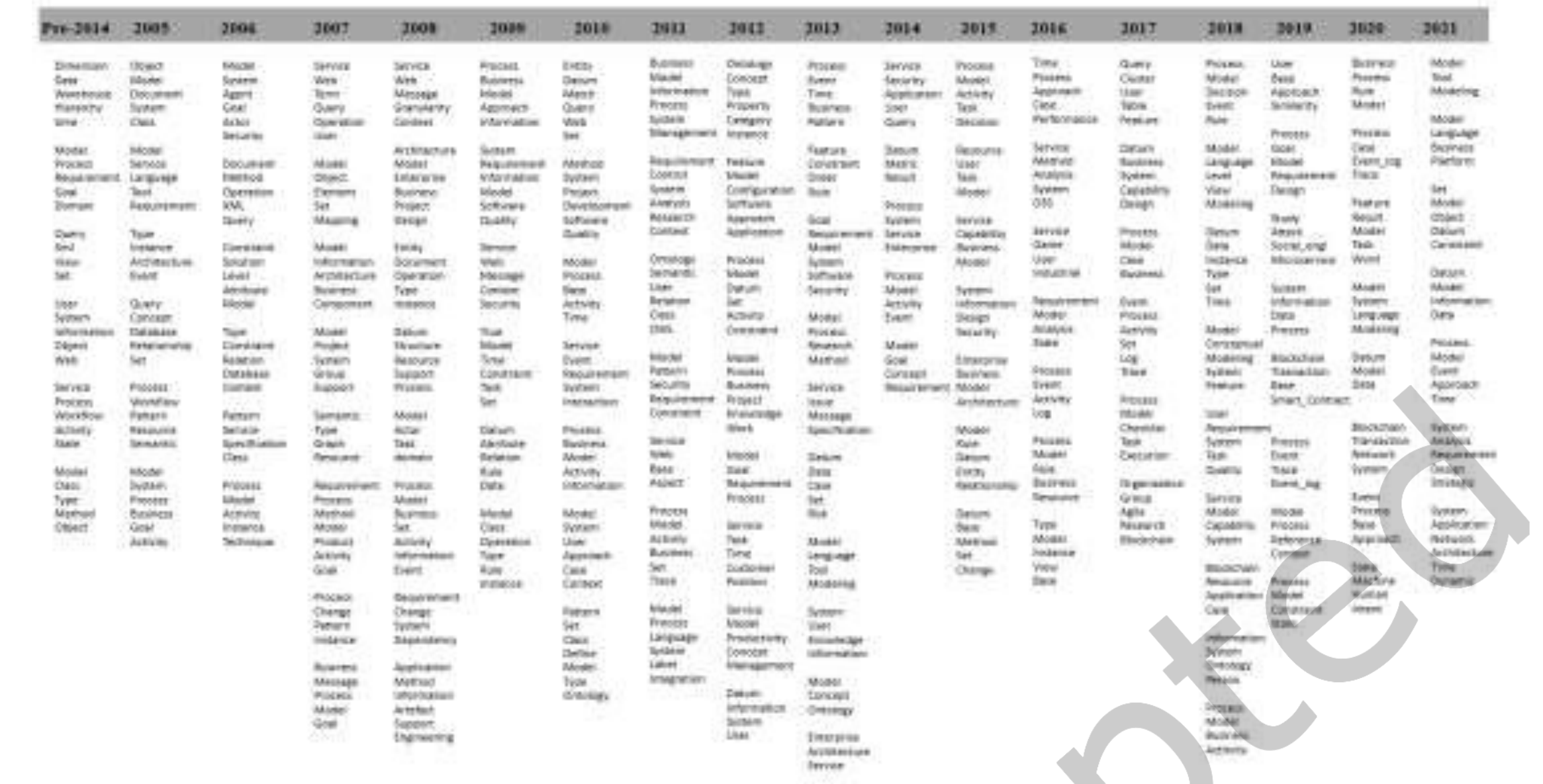

We also performed a topic analysis for CAiSE and EMISAJ separately, similar to that of Table 7. Then, to assess whether these corpora are similar to each other and similar to the initial corpus (Table 5), we used pre-trained language models to measure the similarity among these corpora. We use the sentenceTransformer "stsb-roberta-large," a pre-trained sentence embedding model based on the RoBERTa architecture [133] to calculate the similarity metrics between the different corpora. This is a large-sized model with a large number of parameters trained on a large corpus of text, which has shown to be effective on a wide range of NLP tasks, especially use cases that require a high level of accuracy and understanding of input text [187]. This provides a global indicator of the similarity between the collection of topics across the different corpora in the time analyzed.

The process of using pre-trained language models to compare the similarity between two lists, involves converting the input text into embeddings (i.e., vector representations) and comparing these embeddings between each other to produce a similarity score. The embeddings capture the semantic meaning of the text, with the similarity score being a measure of the similarity between them. The similarity score ranges from 0 to 1. A score closer to 1 indicates a higher similarity between the two topic collections of text; a score closer to 0 indicates lower similarity. The results are summarized in Table 8. Quantitatively, the overlap between CAiSE and EMISAJ with the rest of the corpus is 0.784 indicating substantial similarity [227].

Table 8: Corpora similarity comparisons among the initial corpus, EMISAJ and CAiSE

| **Corpus** | **Initial** | **EMISAJ** | **CAiSE** | **EMISAJ & CAiSE** |
|---|---|---|---|---|
| Initial | 1 | 0.784 | 0.749 | 0.784 |
| EMISAJ | 0.784 | 1 | 0.792 | 0.842 |
| CAiSE | 0.749 | 0.792 | 1 | 0.723 |
| EMISAJ & CAiSE | 0.784 | 0.842 | 0.723 | 1 |

The analysis of CAiSE and EMISAJ versus the rest of the corpus, reveals a substantial overlap between these two and other sources. As can be seen from Table 7 compared to Tables 5 and 6, the topics covered by CAiSE and EMISAJ are very similar with a few specific terms (e.g., set, configuration, artifact, checklist) related to more business processes, compared to Table 6. Still, it reinforces our overall conclusion of the prevalence of business process modeling and related topics in the more recent period of conceptual modeling scholarship. Second, as in the more general corpus, there is a significant interest in representing the structure of the domain with data-oriented conceptual models._We also visually see the common core in Table 9 which shows the word clouds of CAiSE and EMISAJ, with many terms overlapping. This analysis reenforced our earlier findings of the common core in conceptual modeling.

Table 9. Topics from CAiSE and EMISAJ shown as word clouds based on relative topic frequency across all years

| CAiSE Word Cloud | EMISAJ Word Cloud |
|---|---|
| 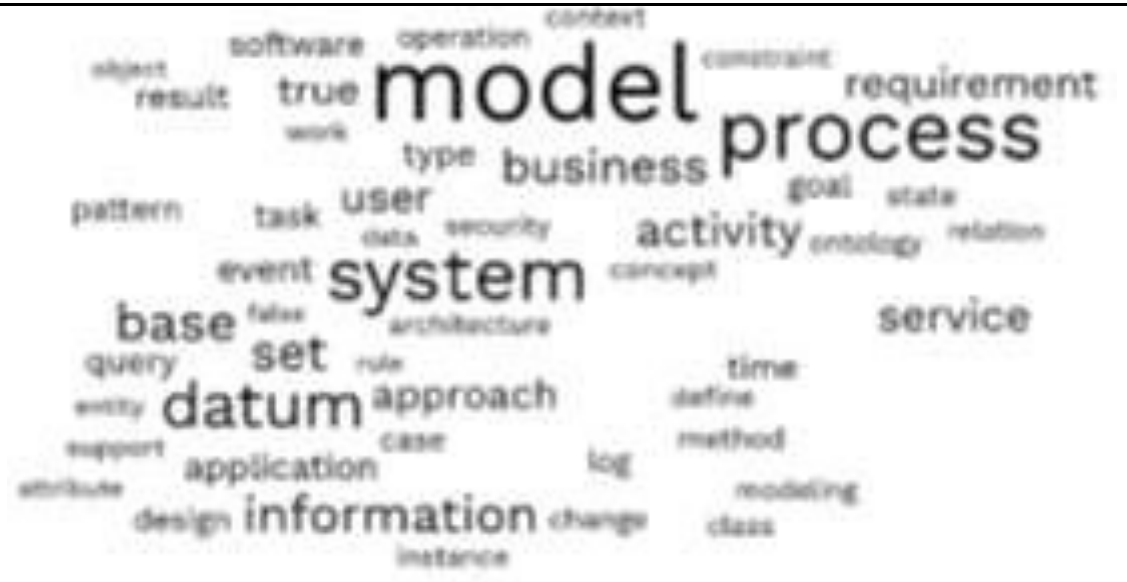 | 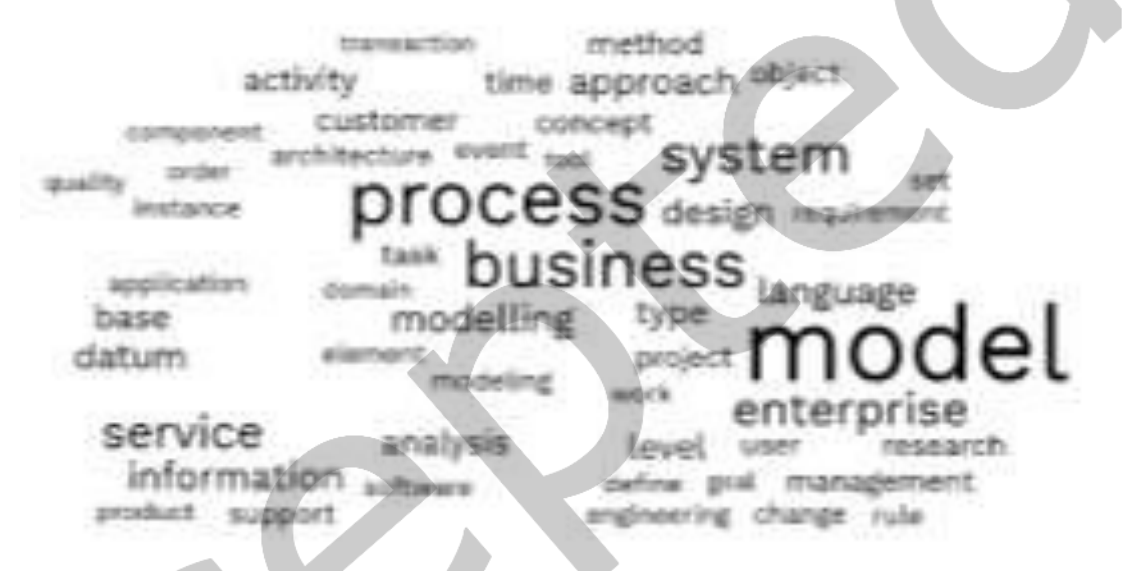 |

*Common core.* We first analyze these two sources relative to the general corpus of 33 journals and conferences. The topic analysis for these two sources together is summarized in Table 7. This analysis is based on the same process and techniques as for the main corpus.

The analysis of CAiSE and EMISAJ versus the rest of the corpus, reveals a substantial overlap between these two and other sources. As can be seen (Table 7 compared to Tables 5 and 6), the topics covered by CAiSE and EMISAJ are very similar with a few specific terms (e.g., set, configuration, artifact, checklist) related to more business processes, compared to Table 6. Still, it reinforces our overall conclusion of the prevalence of business process modeling and related topics in the more recent period of conceptual modeling scholarship. Second, as in the more general corpus, there is a significant interest in representing the structure of the domain with data oriented conceptual models.

*Distinct voices.* The word clouds are generated based on the frequency of the topics between CAiSE and EMISAJ across all of the years of these publications. This offers us a unique opportunity to reveal distinct conceptual modeling perspectives that has been very rarely articulated in conceptual modeling research.

In CAiSE (Table 9, left side) the results depict a community that places conceptual modeling within a broader context of organizational and socio-economic issues. This is underscored by such terms as “users”, “task”, “requirements”, “work”, “operation”, “context.” The software development and database focus are also notable, with terms of “software”, “method”, “application”, “query”, “base”, “set”, and “datum.”

We see a similar pattern of the overlap with the common core in EMISAJ (terms of "process", "model", "modelling"). At the same time, we see EMISAJ's distinct voice. The focus on building business information systems – indeed the terms "business", "transactions", "service", "order", "management" and "customer" are more prominent than in CAiSE or the general (initial) corpus. True to the title of the journal - *Enterprise Modelling and Information Systems Architectures* – we see terms "enterprise", "architecture", "system" and "engineering."

From the exposition of two individual sources, we can vividly observe the distinct voices of different conceptual modeling communities. While sharing interests with other subcommunities, *International Conference on Advanced Information Systems Engineering* and *Enterprise Modelling and Information Systems Architectures* are clearly making their own unique contributions to conceptual modeling. Among other things, these contributions reveal the multifaceted nature of conceptual modeling and the need to better understand and encourage diversity, as well as more holistic approach to modeling issues.

## 3.4 Summary of the results and findings

We can now synthesize the findings across the years to draw general conclusions. Over the last 15 years, we observe a great prevalence of process-oriented conceptual modeling research. Half of the clusters deal with process modeling, of which BPMN is clearly the most popular. After process, the topic of data modeling emerges as the most common. This is interesting and notable. From our analysis of the historic literature (1970s-1990s), data models were the most common conceptual models. In the past 15 years, however, process models began to dominate the conceptual modeling landscape. Several potential explanations exist. First, there is an increased growth of technologies that do not rely on conceptual data models, such as entity relationship diagrams, for their development. These include societally important applications, such as artificial intelligence, natural language processing and machine learning. Very little is known about conceptual modeling within these contexts [72, 135, 184]. There are also notable changes to database storage, including the growth of non-relational databases, such as NoSQL and NewSQL [28]. There is no established approach for modeling these databases using traditional conceptual modeling notations [97, 112].

Under these circumstances, the relative importance of data models continues to decline [11, 136, 196]. At the same time, process models appear to be adapting much better to the new landscape. Indeed, whether using flexible database technologies, or artificial intelligence, the introduction of these technologies in organizations continues to require the understanding of the previous process, as well as the modeling of the new process, even if the details of the new technology are not represented in the process models themselves. Furthermore, the process modeling community has effectively embraced some of these new technologies. For example, active research uses machine learning, natural language processing, and other artificial intelligence-based techniques to mine processes [72, 91, 121, 151]. More broadly, process models demonstrate a high resilience to the changing technological landscape.

Other notable findings include the importance of theoretical and methodological research. This is manifested in the continued development of notations, as well as interest in general ontology. As information technology impact more and more aspects of human existence, it becomes ever more important to develop these technologies based on solid theoretical and methodological foundations. Methodological and ontology work in conceptual modeling is a response to these challenges.

An important observation is that many of the topics of interest to the conceptual modeling community do not occupy a sizable share of the publications or are absent. Examples are the modeling of goals, intentions, dependencies, and contingencies among actors in the domain of modeling, despite persistent recognition of the need to dedicate more attention to these valuable topics [158, 245]. In addition, there is little focus on explosive, emergent technology trends, such as artificial intelligence, analytics, crowdsourcing, blockchain or large language models.

This is a broadening of the focus of conceptual modeling. As Table 6 suggests, research in recent years emphasized the importance of business, organizational applications, context, and modeling capabilities. This is evidenced by terms such as service, risk, execution, business, task, and goal. There is emphasis on evolving techniques and the need to apply appropriate constructs as indicated by the terms graph (for representation), conceptual, class, object, and others. We also compare our results to another large scale literature review on conceptual modeling by Härer and Fill [95], since the methodology, outlets, and years considered, make it appropriate to do so. Table 10 summarizes the comparison and identifies specific overlap in terms extracted by both efforts.

Table 10. Comparison to Results from Härer and Fill [95]

| Topics and terms Härer and Fill [95] | Frequency (this work) | Topics and terms Härer and Fill [95] | Frequency (this work) |
|---|---|---|---|
| **Topic 1** | | **Topic 5** | |
| process | 26 | model | 61 |
| model | 61 | use | 0 |
| business | 13 | case | 4 |
| Service | 8 | software | 1 |
| System | 27 | tool | 2 |
| **Topic 2** | | **Topic 6** | |
| model | 61 | state | 10 |
| language | 8 | event | 6 |
| tool | 2 | model | 61 |
| metamodel | 0 | transition | 1 |
| uml | 15 | system | 27 |
| **Topic 3** | | **Topic 7** | |
| model | 61 | class | 15 |
| transformation | 0 | constraint | 11 |
| rule | 3 | model | 61 |
| graph | 7 | type | 9 |
| element | 2 | object | 10 |
| **Topic 4** | | **Topic 8** | |
| model | 61 | data | 3 |
| system | 27 | schema | 13 |
| test | 0 | database | 9 |
| software | 1 | query | 9 |
| feature | 1 | set | 7 |

## 4 DISCUSSION AND FUTURE RESEARCH

Conceptual modeling emerges as a mature research area that continues to progress in response to the evolving needs of information systems development and use [39]. Conceptual modeling will continue to be used for representation and communication but will be required to adapt to the demands of new technology trends. Figure 5 shows the traditional focus of conceptual modeling and its progression to proposed future research that will consider the emerging technology adaption and use in an increasingly digital world.

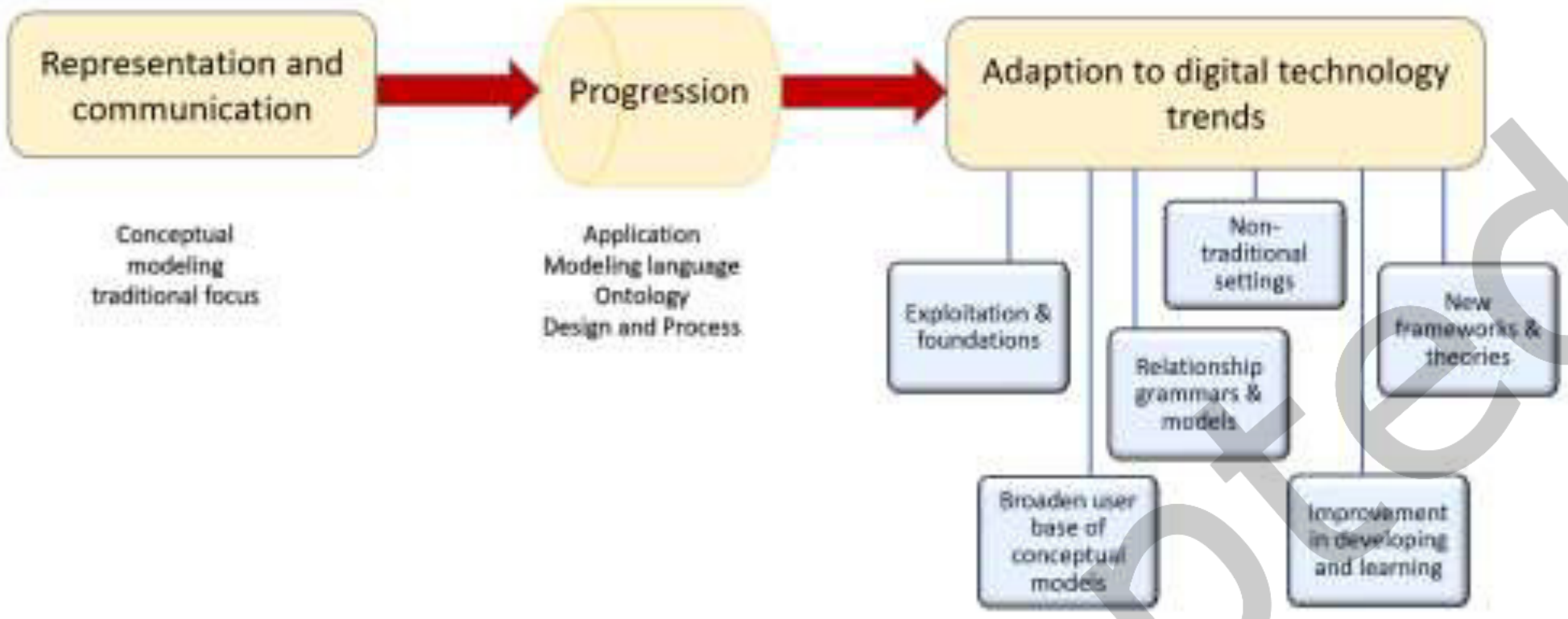

Figure 5. Evolution and continued progression of conceptual modeling

### 4.1 Foundations, Exploitations and Explorations

Over the past 50 years, conceptual modeling scholarship developed a core, as well as periphery areas of research. Our analysis of the literature suggests that stable and persistent themes emerged. We can also identify areas that have, historically, been on the boundaries of what constitutes research in conceptual modeling.

The core themes in conceptual modeling include the development and evaluation of conceptual modeling languages and methods, research on general and domain ontologies, and the application and extension of conceptual modeling languages. Much attention has been on conceptual data modeling and conceptual process modeling; for example, ER model and UML, and BPMN, EPC and DFD, respectively. The analysis of the year 2020 is an example. Less common, but also part of the core of conceptual modeling, are languages that deal with the representation of goals, actors, and values; for example, the i* framework [245]. These languages were mainly oriented towards the collection of user requirements to build organizational information technologies (e.g., ERP), database design, and process improvement. Underlying these conceptual modeling languages is a stream of work seeking to establish strong theoretical foundations for conceptual modeling. Notable theories include a general ontology, such as the BWW (Bunge-Wand-Weber) [235], DOLCE [81] or UFO (Unified Foundational Ontology) [70, 88, 90]. These theories typically assume a materialistic view of reality; that is, that all entities to be modeled are material in nature [215].

Considerably fewer studies have explored topics on the boundaries of conceptual modeling. These include the application of conceptual modeling in contexts where flexible and agile approaches to information systems development did not find utility in explicit and formal conceptual modeling. Such contexts include social media, rapid application development, and web platforms. Despite the lack of widespread use of conceptual modeling in such contexts, further research is needed that is especially tailored to these environments. For example, traditional conceptual modeling approaches appear inadequate for modeling requirements in highly dynamic and heterogeneous online settings, such as citizen science applications [57, 100, 136]. More broadly, a debate emerged over whether conceptual modeling is simply not applicable when developing social media applications, or in agile development, or for modeling NoSQL databases [11, 112, 136]. New conceptual modeling approaches have begun to emerge, suggesting concept modeling is indeed valuable, if not indispensable for these new settings [28, 78, 97, 137, 203]. Likewise, research has noted the limitations of the materialist view of reality [215] and corresponding ontologies (e.g., BWW), which is one of the dominant ontological foundations in conceptual modeling. Yet these ontologies struggle in modeling institutional objects, such as identifiers and social facts [62-64]. Correspondingly, an emerging stream of work focuses on the development of conceptual modeling languages and methods uniquely sensitive to institutional reality [62-64]. Another limitation of a materialist view is that it can be difficult to model psychological intentions and mechanisms, which do not necessarily reduce to the underlying physical processes [140].

The existence of the stable core attests to the healthy cumulative body of research in conceptual modeling, and the relative stability of conceptual modeling as a research discipline. At the same time, to ensure any discipline is adaptive and agile in the face of change, it is important to challenge stable assumptions [6, 120]. Any field of practice needs to encourage both *exploitation* (where dominant ideas are examined and applied) as well as *exploration* (where radically new ideas are proposed and dominant assumptions challenged) [92, 110, 143, 234].

Additional research is needed to deal with topics that are closely related to conceptual modeling, although they might primarily be considered in other areas. One analysis of four dominant conceptual modeling assumptions [184], widely held within conceptual modeling scholarship, suggests that the core in conceptual modeling is deeply entrenched, and that there may not be enough exploration of new ideas and expansion of the periphery of conceptual modeling. These assumptions are that: conceptual models are static representations of physical reality; conceptual modeling diagrams (scripts) represent the deep structure of information systems, which are produced and consumed by humans, and conceptual modeling is an activity undertaken by professional analysts typically for organizational information systems development. Our results support this conclusion because a great majority of the studies focused on what we identified as core conceptual modeling themes. Consequently, a direction for future research is to investigate the extent to which the core of conceptual modeling corresponds to the core of the information systems development and to identify research opportunities that would bring to the forefront themes to address the evolving needs of information technology developers and users.

Conceptual modeling is a diverse and heterogeneous field. Yet, few attempts have been made to take full stock of its diversity, the multiplicity of different perspectives and the value these perspectives provide to the field. We, therefore, conducted two separate analyses for two recognized, additional outlets for conceptual modeling research: the *International Conference on Advanced Information Systems Engineering*

*(CAiSE)* and the open access journal, *Enterprise Modelling and Information Systems Architectures (EMISAJ)*. (See Tables 7, 8, and 9.) This analysis reveals that, overall, these two sources display the common themes in conceptual modeling as identified in our general corpus, as well as their own unique characteristics. The results obtained offer a rich ground for an interesting debate within the conceptual modeling community. As disciplines mature, they develop their own identity. There is a clear sense of identity which has emerged over the years in the conceptual modeling community. At the same time, this identity is not homogeneous. It is, thus, reasonable to conclude that the future of conceptual modeling depends on the ability to work from a set of common, and agreed upon, fundamental concepts and research approaches, while engaging in new, and unique perspectives. This is consistent with how most fields of inquiry progress and mature over time.

## 4.2 Application in Non-traditional Settings

Researchers should continue exploring the typical use cases of conceptual modeling to extend the range of its applications and tasks. New approaches to conceptual modeling are potentially needed due to the: (1) rapid proliferation of open and heterogeneous environments, such as social media; (2) increased system complexity, such as those supporting genomics applications, (3) new non-professional users; (4) rise in unstructured, distributed data and flexible data storage, such as NoSQL databases and data lakes; and (5) new computational opportunities, such as data intensive machine learning and artificial intelligence.

As our analysis shows, there has been a growing effort to apply conceptual modeling in new contexts, such as social media, citizen science, blockchain, big data, robotic automation, artificial intelligence, data analytics, and human genome [33, 89, 94, 122, 137, 141, 160, 174, 214]. Examples of specific topics investigated are: modeling log-based files using a UML variant [181]; automated schema migration and optimization between different NoSQL data stores [46]; artificial intelligence-based approaches to map heterogeneous data models to a relational model in order to take advantage of the ubiquity and maturity of relational databases [133, 246]; just-in-time modeling of flexible data to support varied data analytics activities [35]; JSON schema verification in the context of software interoperability [10, 79]; application of a traditional enterprise modeling language, ArchiMate, to the new context of blockchain systems [48, 229]; conceptual modeling reenforced via augmented reality [157]; multi-level modeling [75, 105, 106]; selecting labels for the diagrammatic elements (e.g., entity types) and natural language processing techniques [224, 225].

As these varied topics demonstrate, the conceptual modeling field is beginning to explore new horizons. Evolving work on the boundaries demonstrates that meaningful and effective conceptual modeling solutions can be formulated in new settings. This suggests there are many unexplored research opportunities for future conceptual modeling scholarship that challenge some of the entrenched assumptions. Practitioners still face uncertainty: does conceptual modeling matter for DevOps practices, social media applications, or other contexts? Questions such as these present an important opportunity for conceptual modeling scholarship to make important, practical contributions and rediscover the value of conceptual modeling beyond its core areas.

### 4.3 New Frameworks and Theories

Both our analysis and consideration of recent publications argue for greater exploration of the applicability of conceptual modeling in different settings. However, extending conceptual modeling to new areas, such as social media, blockchain, or the Internet of things, is not merely a matter of applying traditional conceptual modeling approaches and techniques. Existing research has established that such direct application does not always yield beneficial results. Part of the problem may be that these new settings are dramatically different from the traditional (e.g., corporate, tightly controlled) environments that inspired prevailing theories and frameworks of conceptual modeling. In addition to the development of conceptual modeling languages (or, conceptual modeling grammars), novel frameworks and theoretical foundations may be required.

As trends in diversity and sophistication of information technologies grew, conceptual modeling attempted to adapt. Conceptual data models, for example, carved out an important niche to facilitate systems development and support for the relational database development [206]. However, as the *relative* market share of relational databases began to shrink, so did the relevance of conceptual data models. Relational databases were once dominant for organizational storage, but are increasingly marginal for social media, cloud computing, machine learning, and the Internet of Things applications where NoSQL and NewSQL alternatives dominate. Modeling for social media, cloud, machine learning and Internet of Things remains challenging, with few clearly-established solutions [24, 68, 111]. Similarly, conceptual models were commonly used to support structured development of organizational technologies. However, as more development began to use agile methods, and more applications developed outside organizational boundaries or by end-users themselves [40], conceptual models appeared less relevant.

The lack of broader applications of conceptual modeling to new contexts results from how conceptual modeling is conceptualized. Since most popular conceptual modeling languages were invented long before the age of social media, data-driven artificial intelligence, mobile devices, and virtual reality, they mainly assume a static, unchanging view of domains. As a result, these models are rigid [135] and do not dynamically respond to change.

Second, conceptual modeling has always been understood as a formal, professional activity, better aligned with highly structured development approaches (e.g., waterfall method). Indeed, conceptual modeling was perhaps purposefully ignored by the architects of agile methods [61, 192]. However, conceptual modeling in modern practice is increasingly fluid, flexible, and adaptive [78, 97]. Still, these ideas have not been ingrained in the theory and frameworks of conceptual modeling. The divergence between the ever-increasing demands of the real-world and the capabilities of conceptual models, could result in marginalization of conceptual modeling, both in practice and in research, evidence of which already exists [11, 136, 184].

In response to these challenges, there have been attempts to explore new directions for conceptual modeling [174, 184, 196, 200, 236]. These efforts should, no doubt, continue. For example, there is a lack of understanding of how conceptual modeling may be realized in multiple formats, such as those that use multimedia. Very little work has considered integration of images into conceptual modeling [145] or the use of conceptual modeling in virtual reality [157]. There is no overarching framework that would guide these efforts. There is also a lack a theoretical understanding of when and why the use of more advanced multimedia is needed. Traditionally, multimedia learning theory [147] has been used to justify and better understand the benefits and limitations of using text and graphics [19, 80, 82, 185]. We lack a corresponding theoretical understanding of the use of advanced multimedia for conceptual modeling.

There is a growing body of research that identifies the potential utility of conceptual modeling to support requirements elicitation and development of machine learning models [72, 137, 186]. Here again, we lack a comprehensive framework and theoretical foundations that could anchor these efforts. Hence, there is an opportunity to revisit traditional conceptual modeling foundations when considering new conceptual modeling developments.

### 4.4 Improvement of Process of Developing, Deploying and Learning Conceptual Modeling

With the rapid development of information technologies new opportunities arise related to how conceptual models are created. Traditionally, conceptual models were considered to be diagrams that had to be drawn on paper by analysts. With advances in artificial intelligence, however, including such techniques as machine learning and natural language processing, it is becoming increasingly possible to generate conceptual models automatically or semi-automatically based on a variety of inputs. These inputs can include user documents written in a natural language. Research in this direction has already begun, such as work on process mining from digital traces [66, 91, 149, 150].

There are also efforts to improve the process of conceptual modeling itself. Making conceptual models easier to create as well as deploy is now a new concern. Computer-Aided Software Engineering (CASE) tools became popular in the 1980s, permitting the development of software code based on the semantics captured in conceptual modeling representations [42, 51, 99, 175]. Automated model generation has also been applied within the context of Internet of Things and cloud computing under the "models@runtime" paradigm [18, 36, 50]. Eventually, it might be possible to create artificial intelligent systems that can directly interview potential users as well as consult sources beyond organizational boundaries, such as social media or user-generated images and videos for the generation of conceptual models. Equipped with these inputs, a conceptual modeling design engine may automatically generate a variety of conceptual modeling diagrams. This could be beneficial for ensuring that conceptual models are continuously updated and synchronized with rapidly evolving organizational and social contexts.

Artificial intelligence can also become instrumental for conceptual modeling pedagogy. Ternes et al. [224, 225] have already developed a tool that is supported by natural language processing capabilities and functions as a dynamic assistant for teaching best conceptual modeling practices. The label selection capability of the tool can also be instrumental, even for experienced modelers, because it could potentially compensate for the lack of deep domain expertise. Similarly, future studies could explore the potential of artificial intelligence to improve teaching conceptual modeling, through better personalization aimed at the varied levels of motivation and expertise of the learners.

## 4.5 Broaden User Base of Conceptual Models

As more people become computer literate, and the technology skills of organizational employees continuously expand [114, 162], it is becoming more important to support broad categories of users in their data management needs through conceptual models. This leads to an exciting new research direction for making conceptual modeling understandable and accessible to the masses. More individuals in organizations and beyond are engaging with information technologies and creating their own technology solutions [114, 155, 162]. As Sandkuhl et al. [200] argue, conceptual modeling is due to become a daily activity for everyone. Recker et al. [184] makes this notion more concrete by advancing the notion of citizen modeling.

Correspondingly, research is needed to support these varied users and developers of IT. Consequently, the representations that capture relevant facts about a domain or data stored need to be more accessible than previously required. This questions the fundamental assumption of relying on static, graphical representations as conceptual models. Instead, new formats, such as text narratives should be considered. Alternatively, multimedia videos with highly dynamic and animated scenarios can be used. Some of this work is already emerging (e.g., YouTube videos which animate conceptual modeling concepts[6]).

Traditionally, conceptual models were designed by IT professionals and followed predefined grammatical rules, predicated mainly on abstractions (e.g., identification of classes). However, in many contexts, the representation of concrete objects (instances, entities), can be beneficial, including by allowing those not familiar with abstractions to develop and use conceptual models [102, 137]. In this sense, conceptual models should adapt to the skills, needs, and tasks of users; for example, by leveraging representations based on abstraction (via classes), as well as instances, and narratives.

Samuel et al. [199] challenge the prevailing approach in conceptual modeling to use abstract-based cardinality notations, and demonstrate the advantages of an alternative, instance-based representations. Lukyanenko et al. [135] suggest that in many domains, especially social media, concrete instances, rather than abstract classes should be used to better capture nuanced domain semantics. Eriksson et al. [64] added new refinements to the instance-based conceptual modeling theory by emphasizing the benefits of institutional ontology. Saghafi et al. [197] investigate instance-based representations within the context of query formulation.

Much more research is needed to unfreeze the traditional focus of conceptual modeling research on business development and organizational settings. Because conceptual modeling activities are increasingly becoming mainstream, the research community needs to conduct more scholarship on citizen and grassroots modeling, as modeling by novices, continues to grow.

## 4.6 Relationship between Grammars and Models

As conceptual modeling transitions from an activity solely performed by experienced IT professionals to potentially members of the general public, there is further need to reconsider the fundamental relationship between conceptual modeling grammars and the diagrams (scripts) that are developed using these grammars [30, 236]. Wand and Weber [236] define a script as "the product of the conceptual-modeling process" and suggest "each script is a statement in the language generated by the grammar" (p.364). Hence, traditionally, a conceptual modeling script is viewed as an instance of some already-existing grammar [38, 158, 236]. However, when modeling is undertaken by those who do not know, or are not motivated to comply with, the rules of the grammars, additional consideration is needed for how the script (diagrams) and the rules are related. An opportunity exists to offer a more nuanced, and unified understanding of the relationship between conceptual modeling grammars and the actual conceptual models that are used in practice.

[6] E.g., https://youtu.be/VQGUXm8JjYA

## 5 CONCLUSION

Conceptual modeling is a 50-year-old discipline. It exists as a field that supports the development and use of information technologies. Considering the increasing change in information technology, it is important for the conceptual modeling community to engage in periodic assessment of the progress made in conceptual modeling research and calibrate conceptual modeling scholarship with the developments in information technology practice. This paper examined the progression of conceptual modeling research over the past half century through a collection of over 5,300 papers from relevant journals and conferences. We performed an in-depth analysis of the articles as published in academic sources.

Our analysis shows the on-going evolution of the conceptual modeling discipline. Among the notable findings are the areas of focus, such as conceptual process models, entity relationship modeling, and ontology. While these topics remain of undisputed importance, there are some differences between the capabilities of these conceptual modeling approaches and some emergent technology trends. Specifically, minimal research has been carried out in the areas of artificial intelligence, social media, Internet of things, or blockchain. Likewise, despite interest since the 1990s, relatively less work is being carried out in such interesting conceptual modeling areas as intention, goal, and value social modeling. Yet, the need for more research in these directions is growing, supported by such use cases as modeling ethical behavior in complex artificial intelligent systems or modeling complex adaptive systems. To capture the emerging opportunities, we suggest specific directions for future conceptual modeling scholarship. As information technologies become even more central to human existence, conceptual modeling emerges as vital and incredibly relevant for the digital world.

#### ACKNOWLEDGEMENT

The authors thank the editor-in-chief, associate editor, and reviewers for their insightful comments. We benefited from discussions with our colleagues and the help of research assistants, Elisabeth Do, Yue Shen, Julie Perronnet and Daria Andrievskaya. This research was supported by the J. Mack Robinson College of Business, Georgia State University. We thank the faculty development fund of HEC Montréal, the McIntire School of Commerce, University of Virginia, and the Mason School of Business at William & Mary